# MARKET, ETHICS, AND MORALITY

**Ali Zeytoon-Nejad, Ph.D.**
Associate Teaching Professor of Economics
School of Business
***Wake Forest University***
zeytoosa@wfu.edu

**Abstract:** This paper aims to provide a clear, simplified philosophy of codes of human conduct as they relate to economic and social institutions such as the market and government. It classifies codes of human conduct into two categories: ethics and morals. Ethics are defined as human attributes that aim to 'not make a *negative* impact on others' (aka nonmaleficence), while morals are defined as human attributes that aim to 'make a *positive* impact on others' (aka beneficence). It is explained that ethics comprise a basic, minimal set of norms (e.g., honesty, respecting property rights, and keeping promises) that are clearly indisputable and universally accepted. In contrast, morals consist of a secondary but premium set of standards (e.g., altruism, self-sacrifice, and generosity) that can be controversially subjective and heavily dependent on contextual grounds such as character, culture, and religion. Furthermore, the paper discusses that ethics are connected to the functioning of social institutions, so they are social virtues, while morals are individual virtues, and as such, it is appropriate to delegate the pursuit of personal moral values to individuals, allowing them to pursue them voluntarily at their own personal level. It is also argued that being ethical is a necessary condition for an act to be considered a moral act. Accordingly, it is shown that human acts can be classified under four different categories of "unethical-immoral," "solely ethical," "perceived moral," and "truly moral," and several examples of each type of act are introduced in the paper. Additionally, the paper argues that, unlike its alternative forms of organizing economic activity such as socialism and communism, the free market system only needs to have 'ethical' economic agents operating within it for the market to succeed, but they do not necessarily have to be 'moral' for the market to succeed in accomplishing its mission of improving humans' well-being. Finally, the paper discusses that a 'voluntary' existence of morality (in the sense of helping others out of compassion) in a society can be a great advantage to the performance of any human-based institution including markets (as it can serve as an auxiliary engine to social capital along with ethics and the justice system). However, a 'nondiscretionary', 'forced' incorporation of personal moral principles into the construct of the economic system can be detrimental to the virtuousness of the intended moral act, economic efficiency, individual liberty, individual responsibility, and social cohesion.

**Keywords:** Market, Government, Ethics, Morality, Economic system

**JEL Classification:** A1, A13, H10, P10, P20, P50, P51

## 1- Introduction:

**Moral philosophy**, also known as ethics, is a field of study that explores the nature of right and wrong, good and evil, and the principles that guide human behavior in general. At its core, moral philosophy seeks to provide a framework for understanding the virtuous principles that underpin humans' actions and decisions, and also offers guidance on how humans can lead a good and meaningful life. The great importance of moral philosophy lies in its ability to help people navigate the complexities of the world they inhabit, and to make informed decisions that promote human flourishing and well-being. By engaging in moral philosophy, individuals can cultivate a deeper understanding of their own values and beliefs, and gain insights into the ethical and moral dilemmas they face in their personal and social lives. Thereby, moral philosophy can empower people to make informed individual and social choices and act in such a way that they can contribute to a more just and harmonious society.

**Philosophical terms** around codes of human conduct such as "**ethics**" and "**morality**" are sometimes used **loosely, vaguely, and even interchangeably** in everyday language, which can lead to confusion and misunderstandings about the distinctions between different classes of codes of human conduct. This paper is an attempt to clear up this confusion that apparently exists in public opinion about codes of human conduct as they relate to social institutions in general and to the market as a specific type of social institution in particular. It is argued that there are two distinctive categories of codes of human conduct, called 'ethics' and 'morals' in this paper. It is discussed that ethical codes include human attributes that aim to not cause a negative effect on others, while moral codes include human attributes that aim to make a positive effect on others. While both categories share a common concern with the principles that guide human behavior, it is argued that ethics are social matters, whereas morals are personal matters.

Given these differences, the loose and interchangeable use of the two terms can lead to **ambiguity and miscommunication, particularly in** interdisciplinary contexts such as **economic-philosophical discourse**. Therefore, it is important to be aware of the distinct meanings of ethics and morality in the sense described above and use them appropriately, in order to promote clarity and precision in economic-philosophical discourse. Misunderstanding this subtle distinction between ethical and moral codes of human conduct can turn into a major source of social divide in countries such as the US when it comes to important questions such as what the proper role and desirable scope of the government intervention is in the economy. Comprehending this distinction and removing the fuzziness and unclarity around these two separate principles of human conduct can contribute to a broader consensus in the society about the proper role of the government in the economy and even society as a whole, and thereby increase the acceptance of social order and promote social cohesion within a market-based socio-economic system such as the US.

Accordingly, this paper **aims** to undertake this important mission and shed light on which human attributes are more closely applicable to and essential for the construct of social institutions, and which ones are more closely applicable to individuals and individual responsibilities. By doing so, we can ensure that social institutions, such as the market and government, are able to function effectively towards accomplishing the economic objectives of society, which are, in short, to improve the living standards and well-beings of the citizens at large. In view of this, the present paper takes a philosophical approach to exploring the different natures of different codes of human conduct as they apply to social institutions and attempts to provide answers to the following **questions**: (1) What are different classes of codes of human conduct as they relate to the market and the government? (2) What are the relationships between the two main categories of the codes of human conduct? (3) In what ways do they relate to social institutions such as the economic system? (4) What are the four types of acts that can be defined using the two major categories of human codes of conduct? (5) Which category of human attributes is necessary for the market to succeed? (6) What are the potential consequences of trying to incorporate individual-based moral codes of human conduct mandatorily into the construct of social institutions such as the economic system?

The remainder of the paper is **organized** as follows. The next section provides a formal comparison and contrast for the two major categories of codes of human conduct introduced in this paper. Afterwards, examples of either category are provided. Section 3 discusses the applicability and relevance of these human norms and attributes to the market as a social institution and compares the free-market system with its alternatives in terms of their ethical and moral requirements for them to function properly and succeed in achieving their missions. In Section 4, a conclusion will be drawn from the entire discussion and the main ideas are summarized. Finally, the paper will end with an appendix to provide visuals that depict and illustrate the introduced ideas in this paper in greater detail.

## 2- Ethics vs. Morality:

**Human conduct** is guided by a large set of codes of ethics and morals that dictate appropriate behavior in various contexts. These codes serve as the foundations of civilization and help maintain social order by promoting respect, responsibility, and fairness. They are essential for fostering trust, cooperation, and empathy among individuals and groups, as well as creating a sense of collective purpose and identity. Various sources have contributed to the development of codes of human conduct. **For instance**, one of the earliest known codes **in history** is the Code of Hammurabi, which was created in ancient Babylon around 1750 BCE, which contained laws regulating social relations and economic activities. Other notable examples include religious texts such as the Ten Commandments, the Quran, and the teachings of Buddha and Confucius, as well as philosophical and ethical treatises such as Aristotle's Nicomachean Ethics and Immanuel Kant's Groundwork of the Metaphysics of Morals.

**Codes of human conduct** are often **enshrined in laws**, regulations, and professional codes of conduct, and are reinforced through education, socialization, and cultural norms. They play an important role in guiding human behavior and promoting the **common good**. Codes of human conduct consist of a large set of codes, some of which are more applicable to **social institutions**, while others are more relevant to **individual responsibilities**. Unclarity about such distinctions between codes of human norms can lead to confusion, miscommunication, and even disputes in some social contexts, which can be detrimental to both the virtuousness of human behavior and the obtainability of the common good. Therefore, it is important to be fully aware of the distinct nature of different codes of human conduct as they relate to social institutions.

An easy way of attending to the distinctions between different codes of human conduct is to start with **the Golden Rule**, which is a famous saying in many cultures and religions. In fact, there are several versions of the Golden Rule, including a *negative* or prohibitive version (i.e., **Don’t do unto others what you don’t want done unto you**) and a *positive* or directive version (i.e., **Do unto others as you would have them do unto you**), which are a **negative or positive** ruling governing human conduct, respectively (Flew, 1984). This is a useful basis to start explaining the distinctive essence of different codes of human conduct. Indeed, the comprehensive version of the Golden Rule gives rise to two different sets of codes. The prohibitive version of it simply orders to "**not make a negative impact on others**", while the directive version of it simply orders to "**make a positive impact on others**". In this paper, the former set of norms are called "**ethics**", while the latter set of norms are called "**morals**". Accordingly, there are two types of principles governing human behavior, called ethical principles and moral principal (or, for short, ethics and morals), as per the definitions provided in this paper.

**Both** ethics and morality as two categories of codes of human conduct deal with **distinguishing right from wrong**. This similarity in the general purpose of ethical and moral codes of conduct is probably the reason why the two terms ethics and morality in the sense described above are often used interchangeably by scholars and in public discussions to refer to all codes of human conduct regardless of the inherent differences in their distinctive natures. However, according to the definitions provided in this paper, they have entirely different meanings and address quite distinct human attributes. As such, they should not be confused with each other. Ethics are defined as attributes of 'not harming others', which is also known as the principle of *nonmaleficence* in the literature of philosophy, which refers to refraining from causing deliberate harm towards others. In contrast, morals are defined as attributes of 'benefiting others', which is also known as the principle of *beneficence*, which refers to acts that one should do for the good of another, and includes "acts or personal qualities of mercy, kindness, generosity, and charity" (Beauchamp, 2008).

In essence, **ethics** encapsulate a fundamental and essential assortment of principles, such as honesty, observance of proprietary entitlements, and honoring commitments, which are unambiguously unassailable and hold broad acceptance in all societies, whereas **morals** entail a supplementary yet supreme collection of standards,[1] such as altruism, self-sacrifice, and liberality, which may be subject to contentiousness and heavily influenced by situational factors such as personal dispositions, societal norms, and religious beliefs.

Take the act of **helping a homeless person** as an **example**. The act of giving money to a homeless person is often seen as an altruistic act - an act of beneficence including selflessness and generosity aimed at improving the life of another person. However, the debate surrounding whether giving money to homeless people is truly beneficial to them is a contentious one. On the one hand, **some may argue** that giving money to a homeless person is **not a favor** to them, as it can encourage a vicious cycle of dependency and potentially enable self-destructive behaviors. They may argue that giving money can perpetuate homelessness by enabling individuals to sustain their current lifestyle without seeking necessary support and resources to transition out of homelessness. Additionally, there is concern that giving money to homeless individuals could potentially fund harmful addictions, such as drug and alcohol abuse, which could result in further harm. On the contrary, **others may argue** that giving money to a homeless person is an act of compassion that **can help** to alleviate their immediate suffering, and that giving money to a homeless person can provide them with immediate relief and basic necessities such as food, clothing, and shelter, as well as a sense of dignity and hope, because it implies that they are seen and valued by others.[2]

---

1. As Murphy (2016) puts it, the principle of beneficence is the supreme principle of morality.
2. As another example, one society may deem it virtuous to prioritize the welfare of one's own family over that of their needy neighbors, while another may uphold the act of sacrificing the wellbeing of one's family for the greater good of the community as morally commendable. However, in all societies, it is unequivocally recognized that acts such as dishonesty, theft, deception, providing misleading information, breach

Ultimately, this act of beneficence (i.e., giving money to a homeless person) is highly subject to personal judgement and may align with some people's moral principles and be against other people's moral principles. This is in total contrast with, say, punching someone in the chin for no reason (as an example of inflicting harm on others), which is viewed as a bad, wrong, and unethical act by everyone in every society.

Having defined the two categories of ethical codes and moral codes, **four possible types of human conduct** can be defined, which include "neither ethical, nor moral", "ethical but not moral", "perceived moral", and "both ethical and moral". To illustrate this classification further, **a series of Venn diagrams** are used below and in Appendix 1 in order to depict the domains of codes of human conduct in general, and specify the logical relationships between ethics and morality in that domain in particular. Figure 1 provides a Venn diagram to exhibit the domains of each set of codes. Figure 2 presents a Venn diagram to demonstrate the four types of human conduct introduced above. Figure 3 provides a two-by-two matrix to illustrate the classification of codes of human conduct introduced in this paper in a different way.

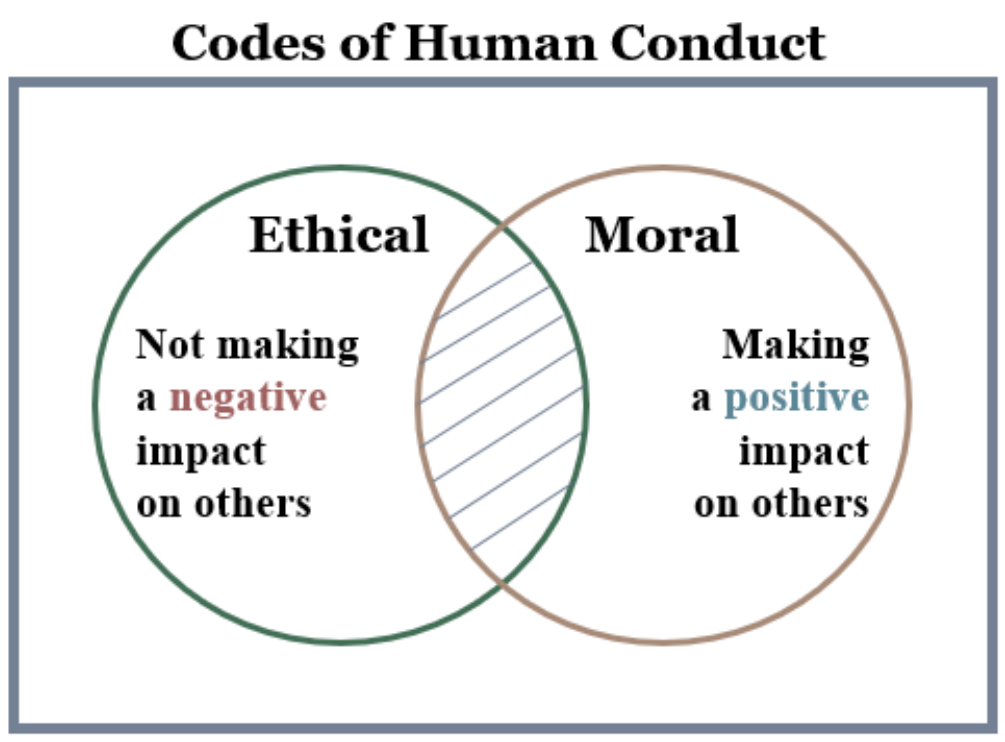

**Figure 1.** Meanings and Domains of Different Types of Codes of Human Conduct

The Venn diagram presented in Figure 1 uses **two overlapping circles** to illustrate the logical relationships between two sets of codes of human conduct so as to show how these two sets of human attributes are similar and different. As shown in Figure 1, the similarity of these two sets of norms lies in the fact that they both fall in the domain of codes of human conduct that guide human behavior in different contexts. This figure also provides the meanings and possible situational relationships between these two sets. As shown above, ethical principles include norms of 'not making a negative impact on others' (aka nonmaleficence), while moral principles are defined as norms of 'making a positive impact on others' (aka beneficence). As demonstrated in the middle of the diagram, there can be some overlapping area between these two domains, which represents the acts that can be both ethical and moral at the same time - which occur when an act does not entail any negative impacts on others, but, at the same time, creates a positive impact on the same group or another group of people. In general, there are **four possibilities for human acts** as they relate to ethics and morals. Figure 2 presents a Venn diagram in which these four possible acts are illustrated and whether these acts are right or wrong in terms of ethical and moral philosophy are shown.

of promises, falsehood, as well as causing harm to one another, whether physical, emotional, or financial in nature, are unequivocally deemed bad, wrong, and unethical. There is a universal agreement about these ethics.

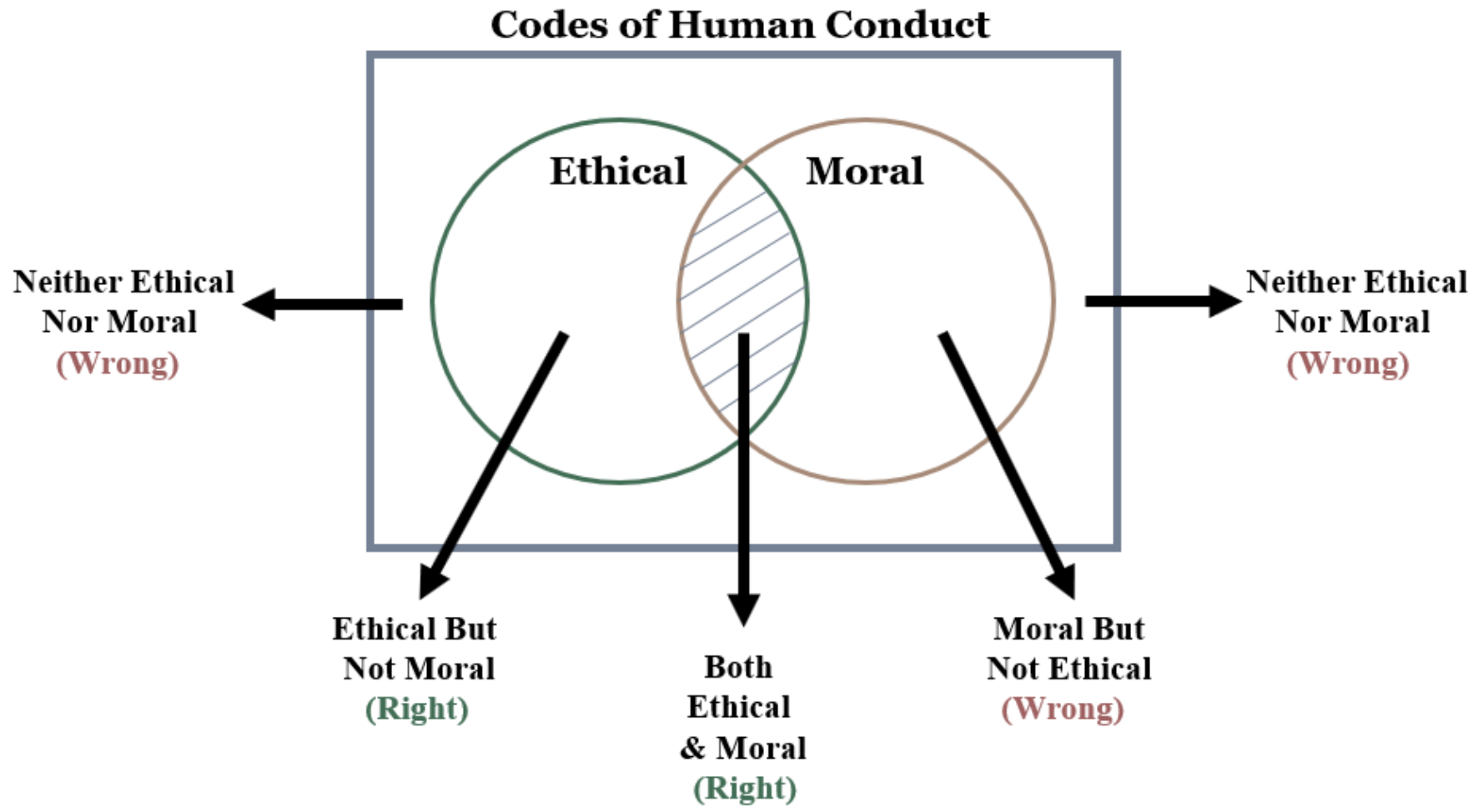

**Figure 2.** Four Different Types of Human Conduct and Their Righteousness in Terms of Ethical and Moral Qualities

Among the four types of acts mentioned above, **two types of acts are straightforward** to understand. First, an act that is **both ethical and moral** at the same time (i.e., an act that does not cause a negative impact on others but causes some positive impact on (some) others) is obviously **desired and right** from the viewpoints of ethical and moral philosophy. On the contrary, an act that is **neither ethical nor moral** is **unambiguously wrong** from the viewpoints of ethical and moral philosophy. Making judgements about the righteousness of these two polar types of acts is relatively easier than the two remaining types of acts, which are discussed below.

A **third type** of possible act in the domain of codes of human conduct is when an act is ethical (i.e., does not inflict harm on others), but may be considered to be not moral (as it does not make a positive impact on others), either. An example of such as act is when someone works ethically and earns a considerable amount of income but opts out of charity activities. This type of act is ethical as it satisfies the basic, essential ethical principles of not inflicting harm on others, and as such, it should be considered right from the standpoint of ethical and moral philosophy.

It is important to note that **a necessary condition for an act to be moral** is for it to **be first ethical**. However, an ethical act does have to be moral in order to be right. As noted earlier, ethics entail a very basic, essential set of codes of conduct. As such, they are the necessary conditions for moral acts to be right. For instance, one cannot steal from others (i.e., violating ethical standards by making a negative impact on others) in order to make donations to a charity organization (i.e., making a positive impact on some others). Such an act becomes immoral in the first place when it violated ethical standards and inflicted harm on the first group. This type of human act should not be called moral, as it violates the very basic ethics, so it is called "**perceived moral**"[3] in this paper, which obviously is a wrongdoing from the standpoint of ethical and moral philosophy. In short, all moral acts must be first ethical (as their first, necessary condition), but not all ethical acts need to be moral.

Put differently and visually, this classification of codes and acts can be framed into a two-by-two matrix of the following form in Figure 3, where the terms on the top and left-hand-side of the matrix represent

[3]. The term "perceived moral" here refers to an act or behavior that is perceived or believed to be morally good or right by the person performing it or by others, but it does not align with universally accepted ethical principles and standards.

the **two types of codes** and the terms inside the four quadrants are the names of the **four types of acts** discussed above. As with the usual form of exhibiting attributes in a two-by-two matrix, the most desirable characteristic falls into the upper-left quadrant (ethical-moral), whereas the least desirable characteristic falls in the lower-right quadrant (unethical-immoral). Appendix 1 presents several additional Venn diagrams and visuals to illustrate these codes, acts, and concepts in greater detail.

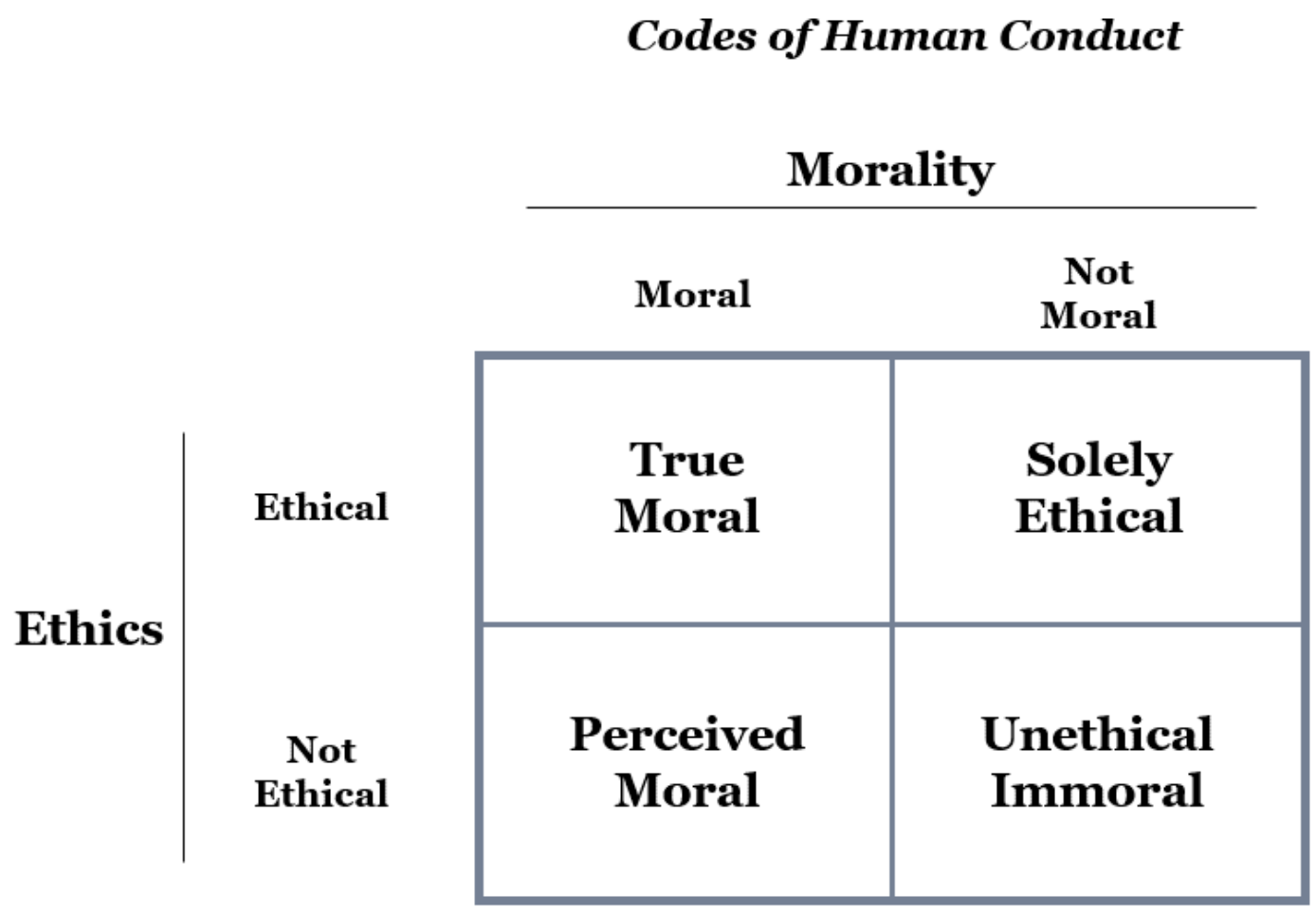

**Figure 3.** Four Types of Acts Classified Based on Two Types of Codes of Human Conduct

To show the distinctions between ethics and morality even more clearly, ethics and morals are contrasted below in terms of several **bullet points** according to the discussion provided in this section.

- Ethics include obligations **to not inflict harm** intentionally on others. Morals include obligations to **create benefit** for others.
- Ethics are a **primary, basic set of codes** of human conduct guiding human beings not to harm others, while morals are a **secondary, but premium set of codes** of human conduct guiding human beings to benefit others. Additionally, being **ethical** is **a necessary condition** for an act to be considered moral, in order to ensure that humans first do no harm, and then do good to others.
- This distinction in the hierarchy of codes of conduct has been implicitly pointed out by several **spiritual leaders and philosophers** as well, such as the Dalai Lama and Arthur Schopenhauer. As **Dalai Lama**, the spiritual leader of the Gelug school of Tibetan Buddhism, puts it, "our prime purpose in this life is to help others. And, if you cannot help them, at least do not hurt them." As another example, according to **Arthur Schopenhauer** (1966) "compassion is the basis of all morality." Arthur Schopenhauer, a German philosopher of the 19th century, posited that compassion is a deep sense of empathy and concern for the suffering of others; however, he says justice is a negative form of compassion. In Schopenhauer's view, rather than actively seeking to benefit others, justice is directed at not harming anyone. In his view, not harming others is a necessary component of a moral society, but it falls short of the more far-reaching, positive form of compassion that seeks to actively alleviate suffering and promote well-being. In Schopenhauer's view, compassion is not always easy to put into practice in organizing the real world that is often characterized by conflict and self-interest.

- Ethics are essential **collective** codes of conduct that society as a whole requires in order to function well and succeed, while morals are more of supreme **individual** codes of conduct that give rise to individual responsibility, and individuals need to follow them to satisfy their internal conscience out of compassion.
- While ethics have to do with the attribute of **nonmaleficence and tolerance**, morals have to do with the attributes of **beneficence and compassion**.
- Ethics, nonmaleficence, and tolerance primarily have to do with the idea that **one must not cause harm to others**, and refrain from acts that are offensive to others, and have the ability to accept the differences of others. In contrast, morality, beneficence, and compassion have to do with the idea that **one should benefit others**, and engage in serving others out of altruism, and have the ability to feel empathy for others' struggles and sufferings.
- Ethics include **universally accepted principles** of conduct, while morals include **individual-based principles** of conduct, and may depend on societal norms, religion, culture, and/or personal values and vices.
- Violating an ethical principle equates to a violation of **fundamental rights** (such as life, individual liberty, and property, which are universally recognized rights), while violating a personal moral principle does not equate to a violation of fundamental rights.
- Put somewhat differently, ethics have more to do with **natural laws**, whereas morals have more to do with **man's laws** and **divine laws**.[4]
- **Examples of** codes of human conduct that fall in the category of **ethics** include honesty, respecting property rights, keeping promises, avoiding causing pain or suffering for others, fulfilling contractual obligations, trustability and reliability, following social norms, work ethics, professionalism, not shirking responsibilities, fairness, and equality in treating people. Examples of violations of ethical principles include telling a lie, murder, incapacitating, stealing, causing pain and offense, defamation, assault, torture, cheating, gaming a system (e.g., fraud, bribe, and collusion), breaking promises, and depriving others of the goods of life and the pursuit of happiness.
- **Examples of** codes of human conduct that fall in the category of **morals** include altruism, self-sacrifice, generosity, charity, mercy, kindness, sympathy, philanthropy, and making sacrifices to empower fellow humans, voluntary time devotions to socially beneficial efforts, emergency aids, and sharing as a way of caring.
- **Ethics** constitutes a basic, essential assortment of codes of human conduct necessary for a social institution to function well. These are social values, codes of conduct, and principles that a social system defines and puts in place to ensure that people will not harm others when pursuing their own self-interest in an interactive social setting. As such, ethics includes **society's beliefs**, which are **homogenous**. In contrast, morality refers to the codes of behavior that an individual, group, or religion follows. Morality is concerned with the rules, principles, and values that an individual adheres to in order to distinguish right from wrong in their own opinion, and is often shaped by

[4]. Natural laws refer to a set of unchanging principles which are regarded as a basis for all human conduct. Natural laws apply to universal human rights and common intrinsic values that are inherent to creation and existence, derived from reason and logic, related to "ethical absolutism," and are not subject to any will, divine, or human principles. In contrast, divine laws apply to religions and are believed to come from God and supernatural beings and are different in different religions. Man's laws apply to the construct of mind and personal beliefs and come from people's personal views of the world and change across different people and cultures and may vary with time and location. As such, both categories of divine laws and man's laws are subject to "moral relativism," unlike natural laws which are eternal and immutable.

cultural, religious, or social norms. As such, morality includes **individuals' beliefs**, which could be **heterogenous**.

- Put somewhat differently, this key difference between ethics and morality can be explained as the fact that **ethics** are seen as more **objective**, while **morals** are more **subjective**. Ethics, on the one hand, involves specific codes of conduct that are widely agreed upon within societies. Morality, on the other hand, involves reasoning about what it means to be a good person, and the virtues that are conducive to that end. This can vary depending on the culture, context, or the individual interpreting the moral principles.
- Ethics is about the **norms of obligation**, while morality is about **moral ideals**.

Developing a comprehensive understanding of these differences between the two sets of codes of conducts and the four types of human acts introduced in this paper are important tenets and fundamental pillars of designing complex social institutions (e.g., the market and the government) in such an ethical and efficient way that can promote social order and unity. The next section delves more into these distinctions as they relate to social institutions such as the market and the government.

## 3- Market, Ethics, and Morality:

This section attempts to distinguish between the two terms "ethics" (defined as **society's beliefs**) and morals (defined as **individuals' beliefs**) as they apply to the proper design of social institutions such the economic system (defined as a combination of the market and the government and their interconnections). Ethics include society's beliefs, which are universally homogenous, while morals include individuals' beliefs, which are typically heterogenous across individuals, cultures, and religions. As explained in the previous sections, while ethics refer to the rules and conducts that are accepted by society, morality refers to the internally accepted standards of individual behavior.

Put simply, **the economic system** is the way a society organizes its scarce resources and economic activities to produce and distribute goods and services. The **market** and the **government** are two main components of an economic system, and they are interrelated in multiple complex ways. The market is a mechanism that coordinates the production and distribution of goods and services through the interactions of buyers and sellers. In a free market system, prices are determined by the forces of supply and demand, and there is some degree of government intervention in the market to deal with the cases where the market fails to accomplish its mission of allocating resources efficiently.

The philosophy of the necessity of government in the economy has primarily to do with the possibility of market failures and its proper scope should be determined by the scope of market failures. The government is responsible for establishing the legal and regulatory framework that governs economic activity and prevents market failures from occurring (*ex-ante*) or fixes and compensates for the undesirable effects of market failures after they have inevitably occurred in the economy (*ex-post*). In fact, the proper role of the government in the economy is trifold: (1) **Establishing the legal and regulatory framework** (through institutions such as laws, regulations, the justice system, and the police to ensure property rights are enforced, contracts are honored, promises are kept, consumers are protected, and the like), (2) **Promoting efficiency in the presence of market failures** (in the case of externalities, market power, economic stabilization, as well as providing and regulating the medium of exchange optimally, protecting common resources, producing public goods that are either not produced at all or not produced sufficiently in the free market including national defense, education, infrastructure, and the like), and (3) **Promoting equality if income distribution is ethically unfair** (i.e., if there is a valid reason to believe that the economic pie in the economy has not been distributed desirably, say, because a group has created a negative externality on another group and have ended up with higher income thereby). In **all these cases**, the government has an **"ethical" duty** to *either* prevent the creation of a negative impact on the people (before the fact) *or* fix and compensate for the negative impact that has already been made inevitably on a group of people (after the fact).

It is also important to note that, since the philosophical underpinnings of the government's role in the economy are rooted in the recognition of the possibility of market failures, the **optimal extent of its intervention** in the economy should be guided by the **scope and magnitude of such market failures** to ensure it does not overreach its necessity. As Thomas Paine (2006) once puts it in his famous book entitled *Common Sense*, **the government, even in its best state, is but a necessary evil**, which has a natural tendency to concentrate power and intervene, so it must be restrained and limited in order to protect individual freedom and economic efficiency.

**The market**, **ethics**, and **morality** are connected aspects that can significantly **impact society's well-being**. As an economic means, a market refers to the structure of buying and selling activities in a particular region or industry. In a market economy, the allocation of goods and services is determined by the interactions of humans as buyers and sellers in a competitive marketplace.[5] **Ethics** (defined as society's beliefs) and **morality** (defined as individuals' beliefs) concern the principles that guide human conduct. In this paper, it is argued that **the market can succeed in accomplishing its mission of improving society's well-being even if humans fail to be moral**. This point is a strong argument for the market's ability to enhance social welfare without a strong necessary condition of all people being moral and having the primary intention of making a positive impact on their fellow humans. It is argued that, for the market to succeed, it is necessary for humans participating in it to be only ethical, which is usually guaranteed by a well-functioning justice system.

The **incorporation of ethics into the socio-economic system** is both necessary and sufficient for the market to function well. This **incorporation of ethics** into the economic system is right both intrinsically and instrumentally as well as justified both principally and consequentially. Building ethics into the market system is required to ensure that the market improves economic efficiency and maximizes social welfare. However**, incorporating morals** (which are subjective in nature in many cases and fundamentally are individual principles and personal matters) into the socio-economic system is both **principally and consequentially wrong**. This occurs, for instance, when the government requires individuals and businesses to engage in charity activities while they are not in a position to be able to do so or even when they have no willingness for such an engagement. It is also **virtuously undesirable** to do so due to the fact that forcing people to engage in activities such as charities involuntarily (either directly or indirectly, say, through price control policies such minimum wage laws) or doing so on their behalf

[5]. The market is a system of exchange that operates according to the laws of supply and demand, with prices and quantities of goods and services determined by the interactions of buyers and sellers in a decentralized manner. The market serves as a mechanism for matching buyers and sellers and determining the value of goods and services somewhere between the Willingness-To-Pay (WTPs) of buyers and Willingness-To-Sells (WTSs) or Marginal Costs (MCs) of sellers. In this system, private individuals and businesses own and operate the means of production and distribution, and decisions about what to produce, how much to produce, and at what price to make transactions are made by individuals pursuing their own self-interest, seeking to maximize their utility or profit through the exchange of goods or services.

without their consent (e.g., by taxing them to finance income transfers to the poor) eliminates the virtuousness of the activity. Additionally, doing so is **ethically wrong** because of the fact that it is a form of an infringement on individual liberty which is one fundamental right. It is also **wrong from a consequentialist point of view** due to the fact that it eventually discourages people from engaging in economic activity in form of production and trade due to the removal of incentives to work hard as a result of income transfers.[6]

For a **market** economic system to function efficiently and accomplish its mission effectively, it **must have a foundation of social ethics integrated into its framework**. However, the inclusion of individual morals into its framework is not necessarily required nor desired, as it may even hinder the system's efficiency and effectiveness by removing the elements of consent, free will, and incentives. It is important to note that **the market does not have anything against morals**, and is not opposed to personal morals *per se*, but rather, the imposition of morals may have adverse consequences on the system's functioning and efficiency. Social institutions such as the market and government should just remain **secular with respect to personal morals** to ensure taking an objective position with respect to subjective matters, and provide an impartial, unbiased playing filed for all people with any personal moral beliefs they may have to be free and have equal opportunity to pursue their own personal, subjective morals, as long as they do not violate universally accepted ethics and do not cause negative external effects on others. Taking this liberal approach with respect to morals provides **a Pareto-efficiency improvement** with regard to the pursuit of personal morals where everyone is made better off without worsening anyone else.

It is also very important to **distinguish** between (1) the **laws and policies that are aimed at guaranteeing ethical behavior** (such as laws related to enforcing property rights and ensuring contracts are honored) and (2) the **laws and policies which are aimed at pursuing subjective moral objectives, which naturally result in restricting individuals' freedom of choice** for those who are not willing to or simply are not in a position where they can engage in the underlying compassion-based activity (such as fiscal policies aimed at implementing compassionate income transfers in the form of, say, Negative Income Tax (NIT) or the Earned Income Tax Credit (EITC)).[7] These are **two separate types of laws and policies** and the government should have **two completely separate stances** on them. The government must strongly support the punishment of crimes and laws preserving property rights but should not set unnecessary laws which have more to do with personal morals. Another type of laws that the government should try to stay away from is those that infringe on individuals' freedom of choice, whether it be about personal matters or economic matters, as enacting these types of laws is not necessary, nor desired. This is because individual choices and private matters fall outside the purview of proper government intervention, and individuals should be free to pursue their own personal morals according to their own

[6]. Such taxation reduces the incentives to engage in production and trade in several ways. It removes incentives among producers as they are not allowed to retain all the proceeds from their production due to taxes. It removes incentives among workers when they see that they are paid the same amount of compensation regardless of whether or not they engage in productive economic activity, and how productively they engage, and whether or not they choose to work effectively. It also reduces incentives for engaging in trade because, as a result of taxing goods for example, the price that buyers need to pay will increase (so fewer buyers will have willingness to pay for the product) and the price that producers will receive will decrease (so fewer sellers will have willingness to sell their products at the lower price), resulting in a reduction in the quantity of trade in the market.

[7]. It is worth noting that this argument is not applicable to transfer-based welfare plans such as Temporary Assistance for Needy Families (TANF), Supplemental Security Income (SSI), and In-kind transfers which are specifically intended for the most vulnerable members of society (i.e., those who are in desperate needs of these programs to survive because of the situations they do not have any control over such as disabilities). This is because the government's support for this group can be viewed as a form of assistance aimed at providing them with the bare minimum needed to survive and recover from their situation. As such, ensuring a basic level of living standards for this group of people should be regarded as a fundamental ethical right, which is essential for preserving human life, which is covered under ethics.

moral virtues, as long as such choices do not cause a negative effect or externality on others.[8] Not only such laws would be an unwarranted violation of personal autonomy, but also such laws would prove difficult to enforce.

**Economic policy** must be focused on promoting economic efficiency and correcting undesired market outcomes due to market failures, such as externalities and market power (e.g., due to monopolies and potential imbalances of power in labor markets, which can make the resulting distribution of income undesirable), which market has not been able to address already through its inherent natural mechanisms. Economic policy **should be objective and neutral with respect to moral values**, which are individual-based in nature. Different people hold different moral philosophies. As an example, fairness can be defined in different ways by different people. If an economic policy is aimed at promoting economic efficiency or correcting a market failure, it is both ethical and moral. Conversely, if an economic policy is set in such a way that it dictates only a single type of moral philosophy, or curtails individuals' freedom of choice, or distorts the market outcome, or even worse, violates humans' fundamental rights, then such an economic policy is neither ethical nor moral.

**One example of such an economic policy** is when the government implements policies that practically coerce individuals to engage in charitable activities (e.g., by a forced transfer of income without any justifiable reasons underlying it - such as an existing market power). Although the objective of the policy seems to be to promote equality (which might seem a moral objective to many), it is a violation of the freedom of choice in the first place (which makes the policy unethical regardless of the result). This is because it is a sort of a forced charity. (Although the author is a huge advocate for charitable activities, but he is an advocate for both charity and freedom of choice.) Charity is a virtuous act if it is pursued voluntarily and free from force or coercion. Rather, **a better economic policy** to achieve the above-mentioned objective may be to provide tax incentives for charity contributions, rather than forced transfer of income. This allows people to have a chance to at least select the preferred form of their charity such that it aligns with their personal moral philosophies.

To continue this discussion and delve deeper into this topic in a more conceptual and structured manner, let us return to the four categories of actions briefly introduced in the preceding section, and elaborate on them as to how they are related to the market, and proceed from there. As previously stated, there are essentially **four types of human conduct** that can be defined based on the two categories of ethics and morals as they relate to the market and government affairs. These acts include:

**(1) Truly Moral Acts (Ethical and Moral Acts):** These include acts that do not make a negative impact on others but generate some positive impacts on (some) others. An example of such an act is when someone earns an amount of money honorably without causing a negative externality on others (i.e., not imposing a negative spillover such as pollution on bystanders) and creates value (intentionally or unintentionally) for others or uses the proceeds of their work to participate in a charitable activity to contribute to promoting the welfare of a group of people, say, the less fortunate. Such an act is obviously desired and right from the viewpoint of ethical and moral philosophy.

[8]. Milton Friedman (2002) cites a memorable quote from a Supreme Court Justice in his renowned book titled *Capitalism and Freedom*, which establishes a boundary for individual liberty. It states that "My freedom to move my fist must be limited by the proximity of your chin." This quote implies that rationality and freedom must be defined and pursued within ethical confines, which in turn is defined as not engaging in dishonest and harmful pursuits.

It is important to note that, according to this definition, a free market is an example of a truly moral means when coupled with a well-functioning justice system. A well-functioning **justice system** guarantees the existence of ethics in the market (in the sense of preventing people from making a negative impact on others) and the notion of **voluntary exchange** guarantees the moral attribute of market (in the sense of making a positive impact on the market participants). This is because no rational person engages in a voluntary exchange unless they are improving their status quo by engaging in the voluntary exchange. Therefore, the fact that voluntary transactions take place at all is a clear indication of improving welfare. As such, such a market is a truly moral mechanism which satisfies both of the required conditions of a truly moral deed (i.e., not making a negative impact on others but making a positive impact on others).

It is important to recognize that economic systems are aimed at **improving the material aspects** of life and not the spiritual aspects of it. Material aspects are objective and easy to measure, but spiritual aspects are subjective and hard to measure. Measuring morality as **a normative concept** and **a subjective criterion** can de challenging and difficult, as its realization depends on the individual's preferences and values; however, the mechanism of free exchange, consent, free will, and voluntary exchange indicates that those individuals having different values **have been impacted positively through the market institution** by engaging in voluntary exchange. Put differently, people would not engage in voluntary exchange if it did not make their lives better, and the mere continuation of voluntary transactions implies that it does. In fact, the free market together with voluntary exchange shows that people have received a positive impact by engaging in free trade, irrespective of what their individual values and personal views on morality are.

**(2) Truly Immoral Acts (Unethical and Immoral Acts):** These acts include those that are neither ethical nor moral. In fact, these include any act that is in **complete contrast** with **the comprehensive version of the Golden Rule**, as discussed earlier. An example of such acts is when someone steals an amount of money from someone and uses it to promote their own well-being without even helping others. Such an act is unambiguously wrong from the viewpoint of ethical and moral philosophy because it is neither ethical nor moral.

**(3) Solely Ethical Acts (Ethical but Not Moral Acts):** Making a judgement about the righteousness of this type of act can be harder and more challenging than the two first types of acts, which were discussed above. These acts include those that do not make a negative impact on others but do not generate positive impacts on others, either, which technically means they are **morally neutral**. An example of such an act is when someone earns a significant amount of money honorably without causing a negative externality on others (i.e., not inflicting harms such as pollution on bystanders) and uses the proceeds of their work to promote their own welfare only and not that of others through, say, a charitable cause. Although such a behavior may not be considered virtuous by some people, it is **still right from the viewpoint of ethical and moral philosophy**, and there is nothing wrong ethically with that type of act (although some may wish they had seen that the individual had also contributed to a good cause for others, too, but it is a personal matter after all).

**(4) Perceived Moral Acts (Unethical but Intended Moral Acts):** Making a judgement about the righteousness of this type of act can also be more difficult than the first two types of acts discussed above. Perceived moral acts include those that are intended to be moral, but they are not truly moral, as they do not satisfy the necessary condition for being a truly moral act, which is to be ethical first. As such, this type of act is called "perceived moral" in this paper. A simple example of such an act is **bank robbery for the sake of charity**, while a thief hero should not be called a hero, since he violates ethical standards by making a negative impact on others in order to be able to make a positive impact on another group of people. Such an act is immoral in the first place when it violated ethical standards and inflicted harm on the first group. This is unambiguously a wrongdoing from the standpoint of ethical and moral philosophy. In short, a stealer-philanthropist is not a philanthropist. He is a stealer and remains there and never becomes a philanthropist.

Another example of perceived morality is "**forced charity**", which is an immoral act as it violates the ethical principles of observing individual liberty and respecting property rights (both of which are examples of humans' fundamental rights). More generally and more philosophically, it is important to note that **the virtuousness of a moral act stems from the voluntary nature of it**, so when charity as a virtuous act is done by force and not consent, it turns into a perceived moral, since it violates the ethical principle of individual liberty when the donor coercively infringe on another individual's freedom and will. It is a perceived moral act (deemed to be moral because of being altruistic eventually), but it is not because it does not satisfy the necessary condition of a moral act, which is to be ethical first.

An additional example of a perceived moral act is when the government **overly and unfairly taxes the rich** beyond the fair extent of their share that they have paid already, which should be equal to their proportionate consumption of the public goods and other necessary services that they have received from the government. However, most governments tax the rich beyond that proportionate share in order to be able to engage in a utilitarian view of morality to transfer income from the rich to the poor, which is a so-called welfare approach that ignores consent as a necessary ethical code to hold for the government to be able to engage in income transfers to the poor.[9] Although the end result may seem to be moral, the tool utilized (i.e., unfair taxes) to achieve that result has violated the ethical principle of individual liberty. As such, promoting income equality through income transfers that are financed by unfair, heavy taxation on a group of people (usually the rich) to promote equality is a perceived moral and not a truly moral act, provided that such a transfer lacks a compelling reason, such as the occurrence of a market failure. This is because it does not satisfy the necessary condition of a truly moral act, and in fact, it violates the ethical principles of individual liberty, people's control over their private property and the justly acquired proceeds of their work. Additionally, it may even be disadvantageous to the poor themselves if such high taxes result in a stagnation in economic activity and the labor market, and thereby cause an increased unemployment rate, which does not serve the initial intention of the policy ultimately.[10]

[9]. One of the main reasons why most governments engage in this type of income transfer is simply because it is politically profitable. That is, when income is taken from a few wealthy individuals and distributed among a larger number of people in the middle and lower class, a politician may make angry a few individuals but makes happy many more individuals, while each of those angry and happy people has only one vote. By doing so, he would lose a few votes from the rich individuals, while gaining many more votes from the middle- and lower-class individuals, making it politically advantageous on net for the politician, who naturally has to pursue strategies that maximize the number of his votes to be able to remain in office.

[10]. In his article "Defending the One Percent," Gregory Mankiw (2013) argues that income inequality is not inherently bad and that policies to reduce inequality may actually harm economic growth. Mankiw suggests that, in most cases, inequality is a natural result of differences in talent, effort, and luck, and that the wealthy play a valuable role in the economy by creating jobs and investing in innovation. He also argues that policies aimed at redistributing wealth, such as progressive taxation, can discourage hard work and innovation, and that a better approach to addressing inequality would be to focus on increasing economic mobility and equal opportunities for all individuals. Overall, Mankiw defends the one percent and suggests that policies aimed at reducing inequality may have unintended consequences that harm the economy as a whole including the poor.

Another example of perceived morality can be **cosmopolitan prioritarianism**[11] or cosmopolitan egalitarianism, especially if it is pursued by non-democratic governments, which do not represent well their citizens' moral beliefs. Cosmopolitan prioritarianism is a moral theory that aims to promote the well-being of all people, regardless of their nationality or place of birth. The theory posits that our moral obligations extend beyond the borders of our own country or community, and that we have a responsibility to work towards improving the lives of those who are less fortunate than ourselves. At its core, cosmopolitan prioritarianism is based on the idea of prioritarianism, which is the view that we should prioritize the well-being of those who are worse off. In a sense, it can be said that it is a liberalist approach to inequality with a global perspective with respect to the political philosophy of income re-distribution. This moral theory implies that when making moral decisions, we should focus on the needs of the most disadvantaged members of society. In a cosmopolitan context, this means that we should prioritize the needs of those who are living in poverty, experiencing discrimination, or facing other forms of disadvantage, regardless of their nationality or place of birth.

Some argue that the theory places too much emphasis on the needs of those who are worse off, at the expense of other ethical considerations. Others argue that the theory is overly idealistic and ignores the complex realities of international politics and economics. **Angus Deaton**, a renowned economist and Nobel laureate in economics, has been **critical of cosmopolitan prioritarianism**.[12] In his book, "The Great Escape: Health, Wealth, and the Origins of Inequality", Deaton (2013) argues that while global poverty is certainly an important problem, we should be cautious about adopting a cosmopolitan approach to addressing it. Deaton argues that a focus on global poverty can distract us from addressing other pressing issues, such as inequality within our own countries. He also suggests that efforts to address global poverty can sometimes do more harm than good, by perpetuating cycles of dependence and undermining local economies. **Instead of focusing on global poverty**, Deaton suggests that governments should focus on creating more inclusive and equitable societies within their own countries. He argues that policies that support education, healthcare, and social mobility can help to reduce inequality and promote greater well-being for all members of society, while also **working towards greater international cooperation and solidarity**.

In fairness, it is also important to note that **wealthier countries can also benefit from engaging in cosmopolitan prioritarianism** and helping poor countries in several possible ways. It can promote global stability (as extreme poverty can lead to social unrest, conflict, and instability), boost economic growth (as it can also stimulate economic growth in poor countries and create new markets and opportunities for trade and investment for wealthier countries), and increase diplomatic influence (as it can enhance international reputation as a compassionate member of the global community and improve international relations). Although there is controversy around the extent of these potential benefits, it is important to recognize that they exist.

---

[11]. Cosmopolitan prioritarianism holds that resources should be distributed in a way that prioritizes the welfare of the worst-off individuals worldwide. As with the liberalism's political philosophy of income re-distribution, it calls for following a maximin criterion on a global scale, suggesting that countries should aim to maximize the well-being of the world's worst-off nations.

[12]. Deaton commences his composition with the following proclamation: "I have thought about and tried to measure global poverty for many years, and this guide has always seemed broadly right. But I currently find myself feeling increasingly unsure about it. Both facts and ethics pose problems."

In contrast, when **cosmopolitan prioritarianism** is pursued by **private entities** (such as The Gates Foundation[13]) based on their own moral beliefs and philosophy, it **leaves no ethical concerns** whatsoever. This is because, in this case, they get to decide on their own behalf and in accordance with their own personal morals, which is fundamentally different from the situation where a government makes those decisions on behalf of its citizens, who may hold a different moral standpoint on cosmopolitan prioritarianism than that of their government.

Although different people may have different opinions about different moral principles, as they are subjective matters in nature, most morals are invaluable spiritual virtues that should be pursued by individuals at an individual level to help their fellow citizens and other human beings and help in fostering their welfare and living standards; however, the point that the present paper is trying to make is that such **moral objectives** should not be set as laws or enforced by the government. This is because these are **"individual" values**. Individual values are virtuous if they are pursued voluntarily and at individuals' will, in their preferred way, and with their consent. The inherent value of a moral deed emanates from the voluntary nature of its execution. Additionally, as Arthur Schopenhauer put it and argued, although "compassion is the basis of all morality", compassion is not easy to put into practice in the real world which is mainly governed by self-interest. As he says, while compassion is a noble ideal, the reality of the world is that it is often governed by self-interest. This makes it **difficult to put moral principles into practice, particularly within social institutions**, and it would be even more challenging to try to enforce them. Given these challenges, it is best to leave the pursuit of moral principles up to individuals as much as possible. While the government can encourage people to be moral and act out of compassion, eventually it should be up to individuals to take responsibility for their actions and strive to make the world a more compassionate place.

In contrast, **universally accepted codes** of conduct **have social aspects** to them, and as such, they must be guaranteed to be obeyed and be in place in a social institution to ensure its success. As such, **integrating ethics into the economic system** in the form of laws and regulations is principally **right** and consequentially advantageous, since it causes social benefit through the promotion of social trust and social capital. However, **trying to build morals,** say, through taxes and income transfers, **into the economic system is principally wrong** (as morals are virtuous when they are conducted voluntarily and freely) and **consequentially disadvantageous**. This is because doing so generates social harm through the deadweight loss of taxes that are imposed for the pursuit of the underlying moral objectives by the government, and removes incentives to participate in productive economic activity (both for the taxpayer and transfer recipient). This results in lowering output, GDP, and income, and thereby the size of "**the economic pie**" will shrink. Moreover, as a result of the degradation of sentiments due to not having control over your own property and proceeds of your work, the size of "**the freedom pie**" will shrink, too.[14] Some legal theorists and scholars such as Pilon (1978 and 1979) believe that the government's authority to regulate behavior is limited only to actions that infringe upon the rights of others and not applicable to those that are to make a subjective positive impact on individuals. This perspective somehow aligns with **John Locke**'s notion of "**natural rights**" as they relate to **ethics** and his notion of "**tolerance**" as it relates to **morals**.

---

[13]. For instance, The Gates Foundation provides global aids to support initiatives in global health, agricultural development, financial inclusion, education, and clean energy in poor countries.

[14]. Milton Friedman has a famous quote which he said as a summary of his views on the relationship between freedom and equality, which are elaborated on in his book "Capitalism and Freedom" (Friedman, 1962). He said **"A society that puts equality before freedom will get neither. A society that puts freedom before equality will get a high degree of both."**

In order to clearly show the distinctions in the proper role of the government with respect to ethics and morals, these two codes of conduct are contrasted below in terms of several bullet points.

- **Ethics** are **social matters and codes** (and objective and universal) and must be ensured by the government as the social planner guaranteeing social order. However, **morals** are **personal matters and percepts** (and subjective and subject to individual values and personal interpretations).
- **Ethics** fall under the category of **social responsibilities** to ensure social trust and strengthen social capital in order to promote social order. **Morals** fall under the category of **individual responsibilities** to help humanity through compassion.
- When it comes to the codes of human conduct, **ethics** constitute the **proper and necessary** scope of the government intervention in the economy, while it is **improper and unnecessary** for the government to intervene in the **individual-moral areas**.
- **Market-related aspects of ethics** encompass a variety of aspects that guide ethical behavior in business and market transactions. Examples of such aspects include honesty (i.e., being truthful and transparent in providing information and full disclosure to ensure informed decision-making)[15], keeping promises (i.e., to ensure that contracts are honored and promised services such as warranty, returns, and refunds are provided), respecting and enforcing property rights (e.g., private property and intellectual property such as copyright and patents through institutions such as courts and the police), observing fundamental rights such as respecting individuals' liberty (i.e., avoiding infringing on people's freedom and instead respecting their consent and the voluntariness of transactions), generating no harms on others (e.g., avoiding imposing negative externalities or spillovers on other parties or bystanders), income equalization when there is an unethical underlying reason behind income disparities (e.g., a negative externality, an imbalance of power in the labor market, or unequal opportunities due to, say, unfair and unequal starting points, unequal rights, or unequal processes)[16], and adhering to the professional behavior that is consistent with one's organizational commitment and obligation as a member of an organization (e.g., adhering to work ethics and performance expectations).
- **Market-related aspects of morality** involve a variety of forms of unnecessary government interference with the natural market mechanisms, individual liberty, or moral principles that violate ethics, or disrupt the natural process of business innerworkings and market transactions, or violate the moral neutrality of the market. Examples of such aspects include imposing heavy taxes to finance the government's engagement in income equalization for the sole purpose of equal outcome (without any regards to the underlying reasons behind the inequality and whether it reflects only unequal levels of effort and effectiveness)[17], forced charity, non-discretionary philanthropy, integrating religious rules into market rules and economic policies (while the

[15]. Honesty is a necessary condition for "perfect information" and "symmetric information", whose absence may lead to market failures, as discussed in economics.

[16]. This type of income transfers and equalization should be called "**compensatory income transfers**" for the unethical underlying reasons mentioned above. This type of income inequality has been produced by an ethical flaw in the market system (i.e., a sort of market failure) or one in the larger structure and context within which the market has operated. Therefore, it is ethical and a social responsibility for social institutions such as the government to compensate for the underlying market failure and eliminate such an inequality through income transfers to fix the effect of the market failure or the ethically flawed context behind the market.

[17]. This type of income transfers and equalization should be called "**compassionate income transfers**", which is done for no good ethical reasons behind it, and is undertaken solely for the purpose of increasing equality out of compassion and altruism, which are examples of subjective, personal, moral objectives, in which a morally impartial government should not get involved.

market and government should remain laic with respect to religions' beliefs), and cosmopolitan egalitarianism.

- Government **involvement in ethical codes increases the size of non-anarchic freedom pie** in society (which ought to be one of the key missions of the government), while the government **involvement in moral codes reduces the size of non-anarchic freedom pie** in society (which contradicts the government's core mission in this regard).
- **Markets succeed** in their mission of serving society and producing better social outcomes **even without morals**, provided that ethics are observed and not violated.
- The government must take **a direct preventative approach** with respect to the violation of universally accepted values (i.e., **ethics**). However, the government can (at most) take **an encouraging approach** (similar to a libertarian-paternalistic, nudging approach) with respect to widely accepted morals in society through indirect policies such as tax deductions for charity contributions, instead of government itself contributing to and getting involved in donations to charity. Such funds at best should be financed through non-tax sources of the government revenue to minimize people's conflicts of interest and potential confrontations in society. Ideally, the government should **remain proactive** with these personal aspects of codes of conduct (i.e., morals). Although the government has the power to intervene, it should not, and instead, it should stay away and remain neutral and secular with respect to these aspects of codes of conduct.
- To ensure that the market does not fail in allocating scarce resources and maximizing social outcome, it is enough for it to operate in an ethical environment, and not necessarily in a moral one. **Examples of market failures** that may arise **due to lack of an ethical environment** include imperfect information[18] (i.e., not disclosing all the available information, as introduced by Friedrich Hayek (1945))[19], asymmetric information[20] (i.e., one party in a market transaction having more information than the other, as introduced by George Akerlof (1970)), negative externalities (e.g., pollution imposed on bystanders), market power (e.g., a monopoly with its typical negative consequences such as higher prices, artificially reduced output, reduced incentives for innovation, and rent-seeking behavior[21]), moral hazard (i.e., changing in behavior

[18]. In economics, **imperfect information** refers to a situation in which buyers or sellers or both in a market have incomplete, inaccurate, or asymmetric information about the product or service being traded, which **can lead to market failures**. It means that some market participants have more or better information than others, which can create unequal bargaining power and result in suboptimal outcomes for some parties. Imperfect information can arise due to various reasons, such as the lack of transparency, high transaction costs, or the complexity of the product or service.

[19]. The idea of "perfect information" in the market was first introduced by the economics Noble laureate Friedrich Hayek in his 1945 article "The Use of Knowledge in Society," which was published in The American Economic Review (1945). (For a more recent published version of his article, see Hayek (2009).) In this article, Hayek argued that the price system in a market economy serves as a mechanism for transmitting dispersed knowledge among market participants, and that the more perfect the information available in the market, the more efficient the allocation of resources will be.

[20]. The concept of **asymmetric information** in the market was first introduced by economist George Akerlof in his 1970 paper "The Market for Lemons: Quality Uncertainty and the Market Mechanism." In this paper, Akerlof, who is also an economics Nobel laureate, discussed how information asymmetry between buyers and sellers can lead to market failure, using the example of the used car market where sellers have more information about the quality of the car than buyers, leading to a market with only low-quality cars, or "lemons," being sold. In fact, asymmetric information is a specific form of imperfect information in which one party in a transaction possesses more information than the other party, and the other party is unable to obtain the information held by the former party.

[21]. **Rent-seeking** behavior refers to actions taken by individuals or firms to gain an economic rent or monopoly power without creating any additional values or benefits for society. It involves the utilization of the political or economic power to gain an unethical, unfair advantage, such as lobbying for regulations that benefit one's own business, manipulating the legal system to gain monopoly power, or using bribery to gain access to resources. Rent-seeking can lead to a misallocation of resources and inefficient market outcomes. As such, it is a form of market failure. It is an unethical act as it makes a negative impact on other market participants because, to gain more surplus and value, it focuses on capturing other participants' consumer or producer surplus rather than creating additional surplus that can be shared fairly among all the market participants.

unethically, say, before and after purchasing an insurance), adverse selection[22], and the principal-agent problem[23].

- **Without ethics** being in place in the market, **markets fail** to achieve socially optimal outcomes on their own. However, **with personal morals** incorporated into the structure of the **market** and the government, they both **fail** economically, politically, and socially, as it reduces the sizes of economic pie, freedom pie, and social cohesion, respectively. This is because morals are typically individual values, and social institutions such as the market and government should be set independent, neutral, and secular with respect to personal morals.
- A strong point of the market system lies in the fact that **markets do not wait for** or rely on **every participant to be moral or have compassion** and good faith in order **for the system to succeed**. Rather, **markets can succeed, even without morals**, as long as participants act ethically and follow certain rules and regulations that prevent market failures from occurring. Any attempt to coercively build subjective morals into the structure of the economic system is very likely to be detrimental to the efficiency of economic systems, even if they are conducted as well-intended efforts in the first place, almost always they end up with unintended consequences that generate undesired results.

**Failure to comprehend the subtle distinctions** between ethics and morals in human conduct can greatly contribute to the fragmentation of society and **exacerbate social divide**, particularly in nations such as the US, where determining the proper role and appropriate extent of government intervention in the economy is a contentious issue. Additionally, this confusion leads to **a lack of clarity regarding the righteousness of the market** in particular and capitalism in general as a market-based economic system. By clarifying these nuanced differentiations and eradicating the ambiguity surrounding these two distinct sets of principles and codes of human behavior, a more comprehensive agreement can be reached in society about the government's proper role, fostering social order, and the acceptance of the market economic system. Unfortunately, this confusion has been one of the primary drivers of social division in the US political landscape, with certain groups benefiting from this obscurity to pursue their own ideology and promote their own personal morals and political interests, some others propagating the unclarity and framing it as a conspiracy or agenda to gain popularity for their so-called discoveries of the conspiracies, and others as the beneficiaries of compassionate income transfers, all of whom have intentionally or unintentionally fueled this division.

The **triumph of free-market capitalism** stems from its ability to **establish ethics as social norms** while **maintaining morality as a personal choice**. This is because free-market capitalism recognizes ethics as objective social values whereas morality as subjective individual values. This permits people to be held accountable when their actions do not align with fundamental rights, and at the same time, ensures to not

[22]. **Adverse selection** is a result of asymmetric information. It is a type of market failure caused by an unethical act when one party in a transaction does not disclose his additional information that the other party does not have, leading to the selection of low-quality goods or high-risk choices. An example of adverse selection occurs when a farmer has a stronger incentive to purchase crop insurance for the parcels of his farmland that he knows have lower quality than his other parcels, and thus are riskier in terms of crop yields, but the insurer is not aware of this information, which results in a negative impact on the insurer.

[23]. **The principal-agent problem** refers to a situation in which an agent acting on behalf of a principal has interests that are not perfectly aligned with those of the principal. It arises due to an unethical act of violating the work ethics of being an agent and not adhering to work ethics and performance expectations, making a negative impact on the principal. It occurs when agents prioritize their own interests over the interests of their principals, resulting in a market failure due to the lack of trust and alignment between the two parties' interests. This can lead to undesirable market outcomes such as moral hazard, where agents take excessive risks knowing that their principals will bear the consequences. The misalignment of incentives between principals and agents can lead to a suboptimal allocation of resources, reduced efficiency, and market failures.

infringe upon the virtuous nature of personal morals by making them compulsory. One of the market's key advantages over other economic systems lies in its ability to attain its mission of enhancing societal well-being with the basic, minimal codes of conduct known as ethics, without mandating or relying upon the more advanced codes of conduct called morals, which are even harder to oversee and enforce. In contrast, the **enforcement of ethics** is easier and more practical, as they **can be guaranteed by a well-functioning justice system**. Free market capitalism entrusts individuals to pursue moral acts on a voluntary basis, making it more preference-compatible[24], more virtuous[25], more efficient[26], and more effective[27].

The renowned economist and Nobel laureate Milton Friedman (2002) once categorized different ways of spending into four classes when discussing taxation in his book "Capitalism and Freedom." These four classes are based on how individuals spend their money. He called them the "four ways of spending money", which include spending your own money on yourself, **spending your own money on someone else** (e.g., charity), spending someone else's money on yourself, and **spending someone else's money on someone else** (e.g., taxation for the sake of income transfers and equalization). According to Friedman, the first two categories (i.e., spending your own money on yourself or on someone else) are the most efficient and effective ways for individuals to allocate their resources because they are driven by market forces and individual choice, so you have a personal stake in the outcome and will make the best choices. On the other hand, **the worst way to spend money**, in his opinion, is spending someone's money on someone else. This is because the person who spends the money has no personal stake in the outcome and may make choices that are not in the best interest of either the person who gave the money or the person who received it. That is, in this case, money is not economized, and the highest value is not necessarily sought. He argues that this type of allocating resources is very likely to result in highest costs and lowest value. **Taxation** for the sake of equalization, according to Friedman, falls under this category of ways of spending, because it involves the government taking someone's money and spending it on someone else, often with little accountability or oversight or even little information about the taxpayers' preferences and moral values and the recipients' preferences and needs.

**Friedman's argument applies** not only to equalization efforts through income transfers, but **also to the whole structure of alternative economic systems** to the market, such as communism and socialism. That the market economic system can achieve the primary objective of the economic system (i.e., enhancing living standards) through the sole implementation of fundamental ethical codes of conduct in the economic environment is in stark contrast to socialism and communism, where ownership is centralized in the state, necessitating employees to be altruistic and productive without incentive, potentially leading to the reduction of publicly produced goods and the destruction of public property.

---

[24]. Once morals are left up to individuals, everyone has the opportunity to pursue the type of **moral activity** (e.g., charity) that they find **most compatible with their personal moral philosophy**. This disaggregation of choice allows for a better matching of moral preferences with moral activities, as opposed to when the government engages in moral activities such as a specific type of charity, which naturally does not match all the society's members' moral preferences for charity (even when the choice made by the government is in accordance with a collective choice based on the majority rule).

[25]. The inherent **virtue** of a moral action arises from the fact that it is carried out **voluntarily**.

[26]. Arthur Okun, a renowned American economist, who is well known for his contributions to the study of income inequality, unemployment, and economic policy, called the tradeoff between economic efficiency and income equality the big tradeoff. Okun (2015) in his book entitled "Equality and efficiency: The big Tradeoff" brings up the **"leaky bucket"** analogy, which suggests that income transfer programs, such as welfare or unemployment benefits, are not efficient because some of the money is lost as it goes through the system, just as water is lost when carried in and poured from a leaky bucket. This means that not all of the intended benefits of income transfers reach the intended recipients. Okun also argued that income transfers have unintended consequences, such as creating disincentives to work or increasing dependency on government assistance. Instead, he advocated for policies that promote economic growth and job creation, as well as measures that reduce inequality and improve access to education and training.

[27]. Charity is more effective than governmental tax-based income-transfer programs in pursuing the goal of equalizing the society because of the fact that charities are pursued voluntarily, and as such, they **do not cause any Deadweight Loss (DWL)**, which is regarded as the social cost of taxation, and is interpreted as loss of gains from trade or loss of market efficiency that a society's welfare experiences due to taxation and contracting the size of real economic activity. Therefore, charity is a more effective tool in pursuing the goal of income equalization.

While communism and socialism heavily rely on public ownership and central planning, their potential success in organizing economic activity efficiently necessitates the people in those societies to be moral and behave morally and always have the well-being of others and helping others in their minds when making their decisions and taking their actions. In contrast, markets do not require such a strong, necessary condition for functioning effectively in society. **People only need to be ethical, not necessarily moral, in order for the market to succeed.** However, being moral and compassionate can add to the social capital of society and contribute to its further success. Competitive markets have self-regulating mechanisms or complementary institutions (e.g., the justice system) that prevent issues that are roots of market failures such as the free-rider problem, the principal-agent problem, conflicts of interests, the problem of externalities, shirking, and moral hazard from arising. Therefore, markets are useful for society without the need for strict moral behavior, unlike other systems of organizing economic activity, where, without people being moral and compassionate, they will fail. **Under other alternative economic systems**, usually morality (in addition to ethics) is thought of as a premise and requirement to exist in the structure and environment of their economic system, in order for it to function well. However, the record of history has typically shown that these assumptions do not hold, and these requirements are not fulfilled in the long run, and those alternative systems perform often very poorly.

In the market system, ethical rules **usually** need to be **enforced by the justice system**. However, **in many cases**, ethical rules are **naturally observed by honorable businesses** in the market system as they are pursuing their own self-interest when business owners become "lifetime" profit maximizers. In a world of perfect information, rational producers would make decisions that maximize their profit in the long run, so they care about their compliance with ethics, their social image, and reputation as well. This focus on ethical behavior is essential for enhancing customer experience, satisfaction, retention, and generating positive word-of-mouth advertising. Under these circumstances, one has to do good to be able to do great. This natural tendency for ethical practice in the world of free press and social media reduces the burden of ensuring ethics on the justice system. In discussing the social responsibility of business, Milton Friedman (1970) once said responsibility of manager of a business is "to conduct the business in accordance with their (i.e., the shareholders') desire, which generally will be to make as much money as possible while conforming to the basic rules of the society, both those embodied in law and those embodied in ethical custom." The term "basic rules of society embodied in law and ethical custom" aligns intimately with the notion of basic, minimal set of norms, which are called ethics in this paper. In a world of free press and social media, it is the ethics, social image, and reputation of businesses that determine businesses' demand, sales, revenue, and profits in the future. Thus, the consideration of lifetime profit to be maximized by firms makes Friedman's statement hold true, and make even the firms that are mainly driven by self-interest incorporate Corporate Social Responsibility (CSR) initiatives that customers care about and value highly into their usual, natural, self-interested operations.

The **market mechanism** (as a process of voluntary exchange) is both ethical and moral, as explained earlier, as long as the **market environment** (as a playing field for market participants) remains ethical. Without ethics, there will be no efficient markets. However, the market and the government (as a social planner to deal with market failures) need to remain neutral with respect to morality, just as the socio-political system must be laic and secular with respect to religion and culture. This is because morality is an individual affair and a private matter, and different people have different moralities depending on their religions, family values, characters, surrounding environment, cultures, personal belief systems, and other such factors. In addition, relating morality to the market through government intervention puts individuals' moral responsibility on the shoulder of the government, and somehow eliminates the individuals' tendencies and obligations to contribute to a moral society. Moreover, probably most importantly, trying to coercively integrate moralities such as equalization of income and mandatory engagement of businesses in charities creates two fundamental problems. First, the virtuousness of the action is removed, and secondly, efficiency gets hurt, which creates a Pareto dis-improvement (i.e., a negative-sum change), which means it makes at least some people in the society worse off compared to the counterfactual (where everyone is free and allowed to pursue their own personal morals as long as they do not harm others), which could have existed in the society if we had not tried to carve only one specific morality into the economic system.

The **confusion** between ethical affairs and moral affairs can have **significant negative effects** on both economic efficiency and social cohesion. This is because it can undermine incentives, fundamental rights, and natural laws, reducing the size of both the "economic pie" and the "freedom pie" in society. To ensure equal treatment of all morals, there should be an absence of personal-moral involvement in civic and governmental affairs. Personal morals and subjective values are divisive, and as such, should have no affinity or influence on the market or government in a free market system. This is a crucial element of enlightenment and modernization that has not yet received the attention it deserves. While both "the secularization of religion from the government and civic environment" and "the separation of personal morals from the government and the market are crucial", the latter has been relatively neglected in both academia and public discourse.

It is imperative to recognize that **personal morals and subjective values possess the potential to be divisive**, and **at best, they only "cluster" individuals** (say, around just a few religions' morals), **failing to truly "unify" them**, and instead result in political polarization. The market and the government should be the guarantor of the free pursuit of personal moral values and not the enforcers of any specific forms of personal-moral percepts. This approach can help to prevent political polarization and promote social unity.

Moreover, the **confusion** between ethics and morals can also **contribute to a lack of clarity about the virtues that markets** and capitalism as a market-based system **are founded on**. Understanding this differentiation and eliminating any mix-up of these two separate sets of codes and principles of human conduct can foster a broader consensus in society regarding the proper role of government in the economy and promote social cohesion and acceptance of social order and market-based socio-economic systems in society. By appreciating the difference between ethics and morals, we can ensure that the market, its missions, and its neutral stance on subjective moral values are understood properly, and the market is allowed to operate freely and efficiently, so that the economic system remains sustainable, just, and prosperous for all, regardless of the subjective moral values they uphold.

It is also important to acknowledge that, **even if** some **morals are** more **widely agreed-upon** in a certain society, they are **not yet part of the necessities for the establishment of social order** and still **remain to be the responsibilities of individuals**. This is because the virtuousness of a moral act stems from its being voluntary. To the extent that the minority size in a population is large, incorporating such a majority's moral principles in the construct of social institutions is wrong (to that same extent). The second detrimental effect of integrating even widely agreed-upon morals into the construct of social institutions is that it creates a false perception among the people over time believing that the government is supposed to be the society's hero and hold responsibility, and they as individuals fail to recognize their responsibilities and start to refuse to accept their own social responsibilities toward their society.

After all, the **"voluntary" existence of widely accepted morals** can be a **benefit** for an economic system that operates within a highly homogeneous society, where almost all the people share a common understanding as for what is the proper way of making a positive impact on others. Then, people in that society would also help each other out of compassion, which can serve as a second engine to strengthen social capital in the society in addition to the primary engine of ethics and the justice system. However, **forcing** individual morals onto the construct of economic system is not only **unethical intrinsically**, but also detrimental to the virtuousness of the intended act, economic efficiency, and individual liberty, as well as dangerous to public unity, societal cohesion, and social order. It can violate the fundamental rights of minorities who believe in different moral virtues.

Additionally, such a **social consensus about subjective morals** is **unachievable** in **highly conglomerate societies** with heterogenous values and virtues such as the US. Therefore, it is better to leave individual morals to be pursued by individuals voluntarily, and not expect the pursuit of a specific set of morals from the market or the government. Taking such **a liberal, tolerant stance on morals is a rational collective choice**, as it creates a Pareto-efficient outcome, allowing all people to pursue their own moral virtues (such as doing charities in the way that they find morally virtuous) without imposing them on others. In this case, they all can improve their social situation (by all groups pursuing their own morals at an individual level) without worsening off others (by requiring a second group to follow the moral principles that the second group does not truly believe in).

The **market** "is" **ethical**, and "must be" ethical, and cannot come to existence without ethics, and cannot survive outside of the ethical domain. This is because people are less inclined to engage in trade, specialization, and work division, say, when they see their property is at the risk of being stolen, or when they believe market participants do not keep their promises. In this case, market participants' interactions would not result in socially efficient decisions under such a high level of uncertainty that the lack of ethics in society may generate. However, the market is not required to be moral. "**People**" should be **moral**.[28] As Milton Friedman (1962) mentioned the market is a machine to run the economy efficiently and should be neutral with respect to personal morals.[29] In other words, the market is a machine to maximize economic efficiency and is not supposed to be a charity organization.[30] As such, it does not

[28]. With regard to social responsibility, Milton Friedman (1962) says "Responsibility is a characteristic of individuals, not of corporations. […] The only entities that can have responsibilities are individuals, not organizations or corporate officials." He also has a famous quote from a speech with a similar theme, where he said "What does it mean to say that government might have a responsibility? Government can't have a responsibility any more than the business can. The only entities which can have responsibilities are people." (Role of Giant Corporations: US Congress Hearings, 1969)

[29]. In his book "Capitalism and Freedom", Friedman (1962) says "Underlying most arguments against the free market is a lack of belief in freedom itself. [...] The market is simply a device for exchanging one set of goods for another. The more clearly this fact is recognized, the less likely are we to confuse the necessity for an effective market with some sort of moral imperative."

[30]. Put simply, the market is aimed at providing an ethical, effective setting in which people are naturally motivated by self-interest to take care of themselves by themselves (as, in this case, they are more likely to take the best decisions for themselves when they see that there is a direct connection between the decisions and efforts they make and the result they will achieve) by focusing on the costs and benefits of their own actions (to ensure those decisions and efforts are promoting the wellbeing of the person making those decisions) while ensuring that they do not make a negative impact on others (by the enforcement of ethics through institutions such as the justice system and government when dealing with market failures). The fact that the free competitive market has two sides (supply and demand), in such a social setup, every seller must serve and

have much to do with subjective, individual morality and should be neutral with respect to heterogeneous personal morals, which are best matched, pursued, and fulfilled when they are left up to individuals.

In sum, **integrating ethical principles** into the market and government affairs fosters economic efficiency, individual liberty, the sense of individual responsibility, and social unity. Conversely, the **coercive introduction of personal moral principles** into the construct of the market and government can impair the intended moral act's virtuousness, decrease economic efficiency, erode individual liberty, undermine individual responsibility, and spark social division.

## 4- Summary and Conclusion:

This paper is an attempt to distinguish between **two primary categories of codes of human conduct**: ethics and morals. Ethics are defined as the set of human attributes that aim to avoid causing harm to others, which is commonly known as the principle of "**non-maleficence**". On the other hand, morals are defined as the set of human attributes that focus on promoting the well-being of others, which is commonly known as the principle of "**beneficence**". The paper argues that being ethical is a necessary condition for an act to moral. It is also shown that human acts can be classified under four different classes of "truly moral," "solely ethical," "perceived moral," and "unethical-immoral," the first two of which are right and the last two are wrong from an ethical and moral standpoint. By clearly distinguishing between these two concepts, we can better understand the various dimensions of human behavior and how they relate to social institutions such as the market and the government.

The paper explains that **ethics** consist of **a set of basic and minimal norms**, such as honesty, respect for property rights, and honoring commitments, which are generally regarded as indisputable and universally accepted. In contrast, **morals** are **a set of secondary but supreme norms**, such as altruism, selflessness, and magnanimity, which may be subjectively interpreted and heavily influenced by contextual factors such as personality, cultural background, and religious beliefs. Furthermore, it is explained that ethics are social norms, whereas morals are personal percepts. As such, **trying to incorporate personal morals into the construct of social institutions** (whether it be the government, the market, the justice system, or the economic system as a whole) **can be divisive**. Instead, the economic system (composed of the market and the government) must be neutral with respect to personal morals. The pursuit of moral percepts should be left to individuals in order for them to be able to attend to their morals on their own, in their own preferred ways, but at an individual level, as long as they do not harm others in their moral pursuits. This stance is somewhat similar to that of John Locke on religious tolerance, who is regarded as the father of liberalism.

satisfy a buyer (and vice versa) in order to be able to make a voluntary transaction with them to generate a surplus (whether it be value or profit) to improve their well-being makes the result of the market also socially optimal. In fact, the pursuit that was initially motivated by self-interest needs to eventually be mutually beneficial to at least one or some other members of the society in order for the former economic agent to be able to receive some surplus, gain, earning, value, or profit out of the pursuit. This stands in contrast to a reliance on a social-planning government, which, as an intermediary, causes a disconnect between people's agency (decisions, authority, or efforts) and their well-being. Conversely, the market gives both authority and accountability to the individual to ensure that they either take care of themselves or bear the consequences of their actions or inactions by themselves. When individuals have control over their own lives and get to receive the proceeds of their efforts directly, they get to realize that they will bear the consequences of their actions and accept their own failures. No government can perform miracles for its citizens. Entrusting the fate of the people to the government gives rise to the principal-agent problem, which economic history has proven to cause more harm than benefit.

Accordingly, to formulate **a logical, systemic reasoning for** the importance of **moral tolerance and neutrality** of social institutions with respect to personal morals, the **three following arguments** can be provided: **(1)** The government and human beings cannot reliably evaluate and consensually conclude the righteousness of alternative subjective moral standpoints. **(2)** Even if such a consensus were achievable, enforcing a single type of moral to an entire population would not yield the intended desired results, because moral beliefs cannot be coerced through violent means. **(3)** Even if attempted, coercing moral homogeneity would lead to more social divide, unrest, and disorder than allowing for moral heterogeneity and embracing diversity.[31]

Therefore, it is important to recognize and appreciate moral diversity and heterogeneity, which can contribute to a **more inclusive** and **tolerant society**. Just as with religion, as an example of a personal matter, which should be kept separate from the government's affairs, **personal morals should be kept separate from the affairs of the market and economic system** in a broader sense. The ability to distinguish between these two codes of human conduct is extremely important in defining a just, fair, and efficient socio-politico-economic system. As an example, this distinction can shed light on what the proper scope of government intervention is in the economy. According to the classification of the codes of conduct in this paper, the three following statements can be inferred with respect to the role of the government in regard to, say, income equalization in an economy:

- Government (as a modifier) **must equalize** income to the extent of income disparity that is deemed to be due to unethical sources, such as an imbalance of power in labor market or imposing a negative externality by one group to another group.
- Government **must not equalize**/transfer income to balance out outcome while disregarding how much effort, time, risk, expertise, and other factors one has expended in earning that income.
- Government **must not transfer** income for the sole purpose of promoting equality out of a sense of compassion. This is a clear example of a "perceived moral act", as it is unethical to forcibly take away one's earnings to help someone else (even with an altruistic rationale and intention behind). Instead, such altruistic decisions should be taken at an individual level, with people being free to give to charitable organizations and Non-Governmental Organizations (NGOs) as they choose. This approach ensures that individuals are able to exercise their free will while promoting freedom and preserving the virtuousness of the act.[32]

---

[31] When the government takes **a neutral position** towards personal morals, the diversity of morals turns into an **opportunity** for society and strengthens social order. When the government takes **a biased position** towards only one or a few specific morals, the diversity of morals can turn into a **threat** to social order.

[32]. An important consideration that must be taken into account with regard to this argument is that whether the **outcome of the market mechanism** is ethical is contingent on the ethicality of the **initial distribution of wealth** (i.e., the accumulation of income in the first place). In other words, **the allocative outcome of the market mechanism is as ethical as the initial distribution of wealth entering the mechanism is ethical**. The status of equal opportunity is sometimes influenced by the ethical degree to which wealth is distributed within a society. If this distribution is unfair due to an unethical event in history, such as certain forms of biases against minority groups, then compensatory mechanisms must be put in place by the government to empower the impacted generation who have been impacted directly or the subsequent generations who have been impacted indirectly by the propagation of the impact of the unethical event in history into their lives. Both groups have been disadvantaged by that unethical event in history, and thereby both groups should be compensated through compensatory mechanisms that most effectively empower the disadvantaged and enable them to catch up until the effect of the adverse, unethical event has phased out from their lives, and all groups have equal opportunities in the market system.

However, such **compensatory support programs should not be intended to be perpetual**. Rather, they should continue only until the time when it is believed that the adverse effect of the unethical event has phased out from the fate of the disadvantaged over time, and that all groups who put in equal efforts are deemed to have equal opportunities in the market system. An example of the reasonable duration of such compensatory mechanisms is the Supreme Justice Sandra Day O'Connor's opinion in the case Grutter v. Bollinger (2003), where she wrote an opinion on affirmative action and allowed for **a phasing-out period** for the use of race as a factor in admission decisions. In this case, the Court upheld the University of Michigan Law School's use of race as a factor in admissions decisions to promote diversity, but also noted that the use of race should be phased out over time as society becomes more equal and race becomes less of a defining factor for success.

This type of **categorization of codes of human conduct** with respect to the role of the government and the market can **eliminate many grounds of division in societies**, and instead promote social unity through the development of a clear, comprehensive, and mutual understanding of inherent differences in the essence of codes of human conduct. The government as a social institution should s**olely attend to social issues that cannot be solved by the market or the society itself**. It is not appropriate for the government to attend to personal morals. The government should attend only to ethics that have social aspects to them. Imposing personal morals on all people in an entire population and forcing them all to think the same as one person or a group of people do is unethical.

As Adam Smith once put it, “It is not from the benevolence [or compassion or morals] of the butcher, the brewer, or the baker, that we expect our dinner, but from their regard to their own interest.” That is to say that when the self-interest inherent in humans is naturally pursued by people within an ethical framework (and not necessarily a moral one, which requires a stronger condition), in which no one is allowed to cause a negative externality on others, the simple **combination** of **self-interest** and **ethics** together will satisfy all the requirements needed to turn a society into **a thriving society**.

That the market will succeed even with a **minimal** set of codes of human conduct (i.e., ethics) in place is a strong point of the market compared to its alternatives, while other alternative systems require a **maximal**, comprehensive set of codes (i.e., morals) in place to guarantee success. Such a maximal, comprehensive set of codes are **impossible** to exist or **continue to exist** in a society where **self-interest ultimately prevails**, and **flawed mechanisms** due to individual incentives including moral hazard, the free-rider problem, shirking, the principal-agent problem, and the like emerge sooner or later.

The fulfillment of the market’s primary goal, namely the betterment of the quality of life for the masses, hinges on the establishment of **a sturdy ethical framework**. When market activity, defined as a series of self-interest-based transactions, is executed under the guidance of an impeccable justice system, it transforms into a powerful ethical and moral tool. The **justice system** safeguards its **ethical identity**, whilst the essence of **voluntary exchanges** ensures its **moral ascendancy**. One can infer that any voluntary transaction is bound to enhance the welfare of the participants, for no reasonable individual would willingly enter an exchange with the sole intention of diminishing their own well-being. The very existence of voluntary transactions attests to the progressive nature of commerce, for those engaged in such transactions would invariably opt for an improvement in their situation over retaining the status quo.

It is crucial that **ethics are ingrained in the economic system**. Only then does the market as a mechanism become moral, which means it is not only ethical but also makes a positive impact on those who participate in it. However, the market is not supposed to and cannot have all morals of all individuals built into it. The main mission of the government with respect to the market is to ensure that people remain ethical and do not make negative impacts on each other as they are interacting within markets, which, in short, means market failures do not arise.[33] However, it is **not within the purview of the**

---

Over time, **the market provides equal opportunity which facilitates social mobility**, enabling those who put in the required effort and work hard to change their social class and climb the ladder of success, regardless of where in the social hierarchy they started. As a result, subsequent generations will begin at higher levels on the social hierarchy than their preceding counterparts, and the effect of the unethical historical event will phase out and eventually disappear from the system. In the end, it is important to note that even if the disadvantaged groups are not provided by their society or political power with the compensatory mechanism that they deserve to catch up faster, they should not go against the market mechanism. This is because the market mechanism still provides them with the best opportunity for a degree of social mobility that is much stronger than that provided by any other economic systems.

[33]. **Different cases of market failures** can be attributed to **different forms of ethical violations**. Take negative externality, market power, and imperfect information as three prime examples, in which case some market participants cause a negative impact on another member of the society

**government to dictate which subjective moral principles people ought to be compelled to abide by** in order to make positive impacts on each other, which are referred to as morals in this paper.[34]

**Society** must be kept ethical by the government and justice system to function well, but **individuals** should strive to be moral on their own to conform to their own conscience effectively (by observing their own moral values). While it is not the government's role to dictate subjective moral values, in highly homogeneous societies where there are widely accepted morals on which a very large majority of the society agrees upon, the government may engage in the encouragement and dissemination of such morals through synergic efforts or libertarian paternalistic approaches such as nudging that are fair and treat all morals equally.[35] Nonetheless, measuring such a fairness and equal treatment can be challenging.

In this respect, there are **several considerations** that can be taken into account. First, the government's engagement in the encouragement of morals in **societies that are morally homogenous** to a great extent still leaves some minorities morally unattended. The larger the size of the minority in such a society, the more problematic the government's engagement in subjective morals in that society. The government's engagement in morals (say, in highly homogenous societies in terms of their morals) **should be limited to efforts and mechanisms that do not entail significant cash outlays or transfers** to ensure they remain least problematic ethically. An example of such attempts can be the government only using its **synergic capacity** to help private entities to contribute to a good cause that is consistent with the majority's morals. As other forms of the government engagement in morals, the government can also incentivize moral initiatives through measures such as **tax holidays** and **tax credits**. While these approaches are not without their challenges, they may be less problematic to citizens than direct government spending on moral causes. Additionally, **regulatory processes** to attend to morals tend to create less opposition or resistance to moral-based policies than re-distributive processes.

Moreover, the level of sensitivity to the incorporation of morals into a country's market-government structure may also depend on factors such as the nature of its economy and the degree of political polarization. For instance, **nations that heavily rely on non-tax sources like natural resources (e.g., oil-based economies) to finance their government budgets** may be less sensitive to the coercive incorporation of morals into their market-government structures.[36] This could be due to the fact that such re-distributive efforts are not funded by taxpayers' money, but instead by the revenue from the sales of natural resources, which creates a disconnection from (or a weaker connection to) the concerns of the general public. This disconnection creates less sensitivity to moral-based, compassionate re-distributive efforts pursued by the government. Similarly, **democratic countries with a strong majority of people believing in similar subjective morals** may also be less sensitive to moral-based, compassionate re-distributive efforts pursued by their governments, as evidenced by the effective implementation of such policies in Scandinavian countries. Conversely, **politically polarized countries with a rather uniform**

---

by, say, pollution, pricing power and trade reduction, and withholding information, respectively. Other such examples could be the problems of the free rider, the tragedy of the commons, the so-called "moral" hazard (although in our definitions here it is in fact an "ethical" hazard), shirking, and the principal-agent problem. These cases all result from the violations of different ethical codes, either societal codes of ethics or professional codes of ethics. In fact, none of these have to do with the moral principles of making a positive impact on others out of a sense of compassion.

[34]. According to this definition of the proper scope of government intervention in the economy, "compensatory income transfers" for the unethical underlying reasons occurring behind them fall in the proper scope, while "compassionate income transfers" solely for the purpose of increasing equality out of a sense of compassion and altruism falls out of the proper scope of government intervention in the economy.

[35]. Such efforts should be designed in such a way that they do not require any or much of the tax revenue collected from the group of people who may have different morals than those the government pursue in its compassionate programs.

[36]. It is important to note that such a revenue-spending structure of the governance in an economy tends to reduce the government's accountability to citizens and may pose a threat to democracy as well.

**distribution of political-ideological disparities** are more sensitive to the incorporation of subjective morals into their social systems, as seen in countries where the political views are almost equally divided (such as politically polar countries with an almost 50%-50% composition of political views). This is vital to recognize that personal morals and subjective values could be **divisive**, and **at best, they tend to "cluster" people** rather than unify them, which can result in political polarization and social division. Therefore, the government's engagement with subjective morals should be avoided, and if pursued under some circumstances, it must be balanced and based on principles of **fairness** and **equal treatment** to all to the extent possible.

In sum, incorporating ethical principles into the market and government unifies society, while introducing subjective moral principles can lead to a social division. Although the **voluntary existence** of widely accepted morals **can be an advantage** for an economic system in the form of strengthened social capital, a **coercive integration** of individual morals into the construct of economic system is not only intrinsically unethical, but it can also cause many potential **problems** and **threats** to the society. In particular, incorporating ethical principles into the construct of the market and government leads to promoting economic efficiency, individual liberty, individual sense of responsibility, and social unification. Conversely, the introduction of (personal) moral principles into the market and government can have adverse effects such as the loss of the virtuousness of the intended moral act, economic inefficiency, loss of individual liberty, diminishing the sense of individual responsibility, creating social divisions, and if taken to extreme measures, even posing a threat to social order.[37]

[37]. Finally, it is important to re-iterate that the author has nothing against morality and compassion. Instead, he is a huge advocate for compassion and sympathy, and encourages each and every one of us individually to try to engage in the pursuit of these goals to the greatest extent possible. However, the whole point of this paper is that such decisions are personal matters, and coercively incorporating them into the economic system is more costly than being beneficial for the society as a whole. Another major point that the paper makes is that the market will succeed in accomplishing its mission for the society even when people are not moral in their decisions. This is a strong point for the market that it needs just a very basic ethical environment to be able to serve the masses at large. Other than that, the author's personal view on morality at the individual level is the same as that of an Indian spiritual leader and a motivational speaker called Gyanvatsal Swami, who once said: "There is a big difference between **growth, progress, and success**. If you increase the turnover of your company, it is called '**growth**', not success. Increasing your materialistic possession is your growth. That growth, if it is aided by ethics, discipline, honesty, and norms, it is called '**progress**'. So, growth plus ethics is progress, and that progress plus humanity, morality and spirituality is called '**success**'."

## Appendix: Visuals Illustrating the Ideas Introduced in the Paper

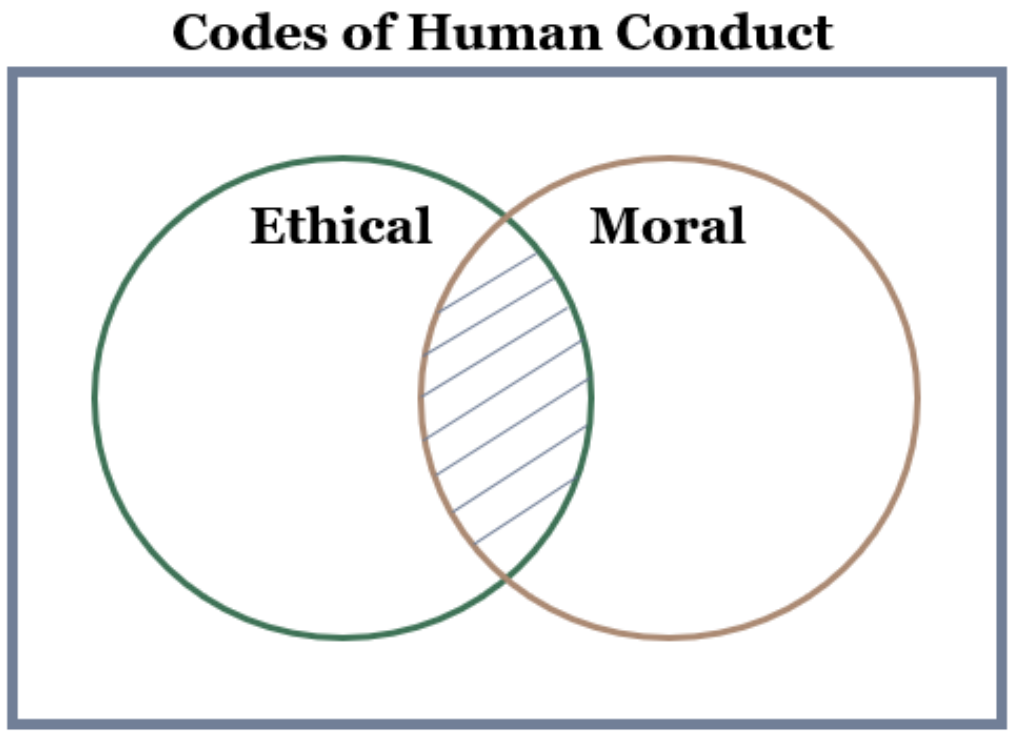

**Figure A.1.** Different Types of Codes of Human Conduct

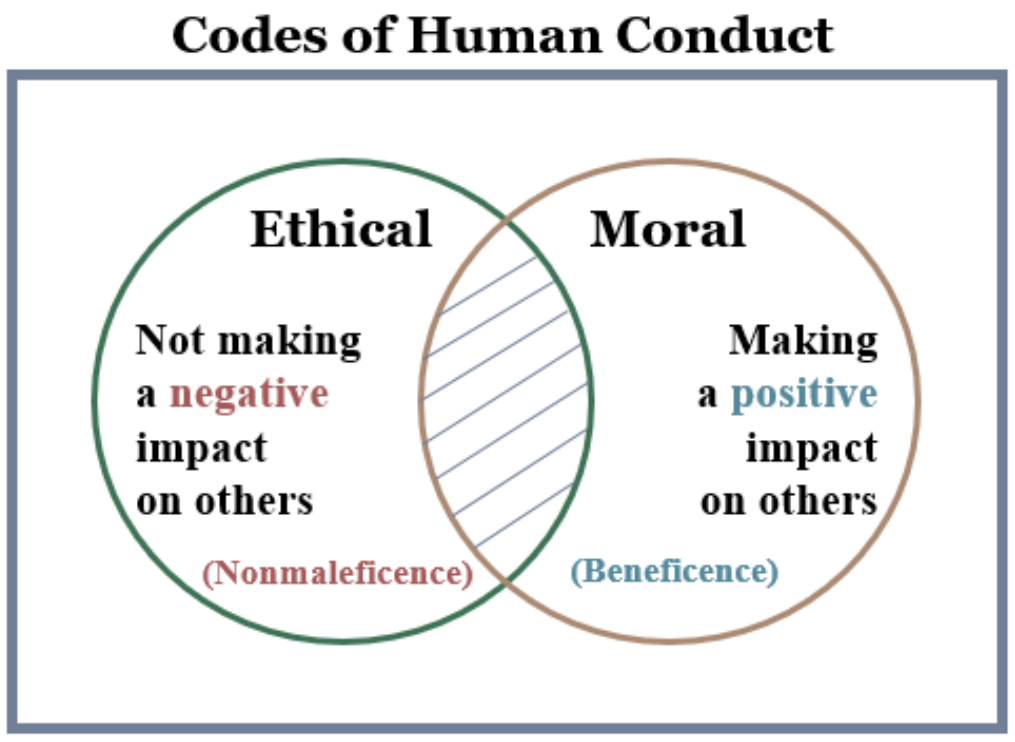

**Figure A.2.** Meanings of Different Types of Codes of Human Conduct

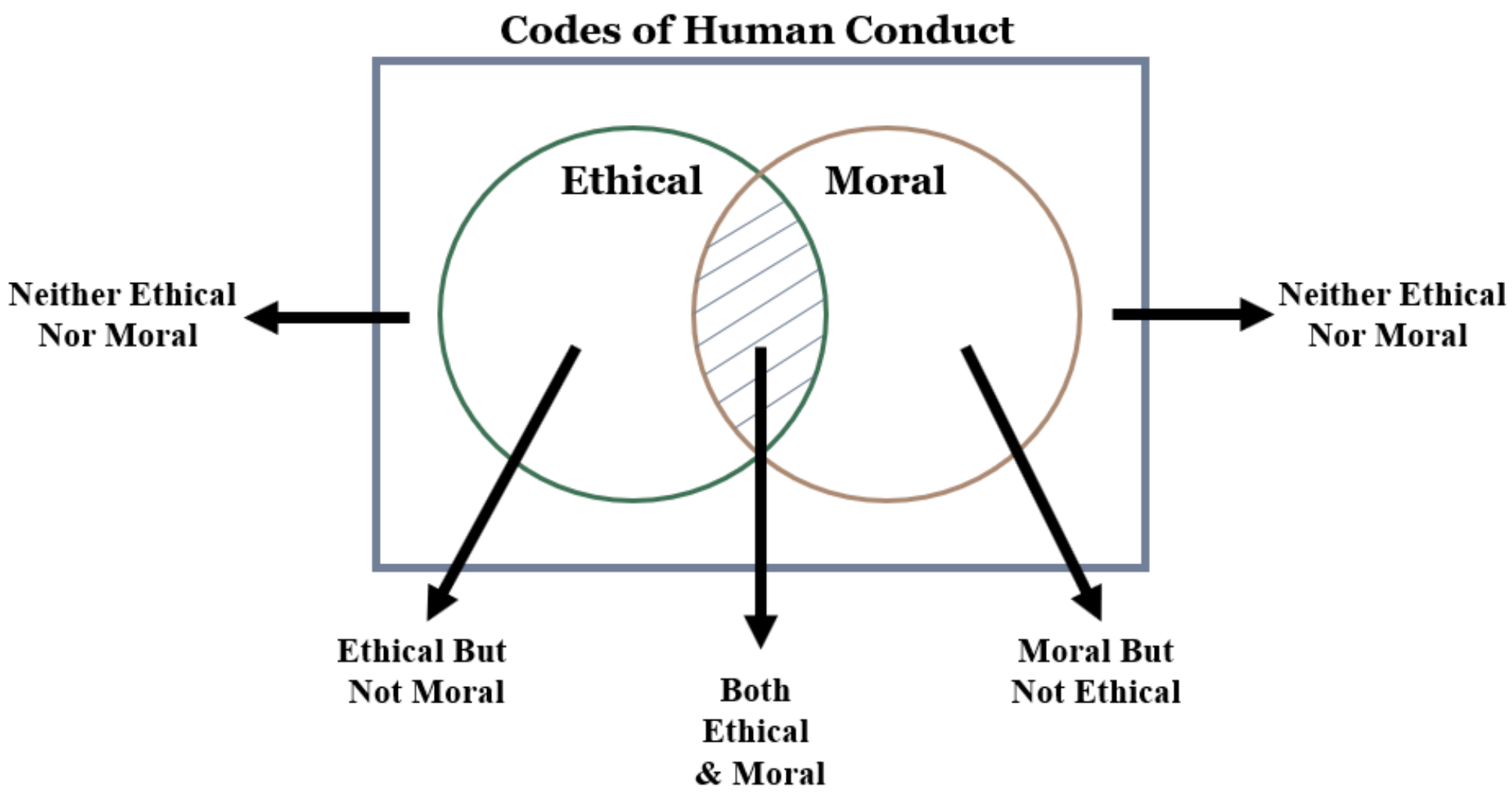

**Figure A.3.** Meanings of Different Types of Codes of Human Conduct

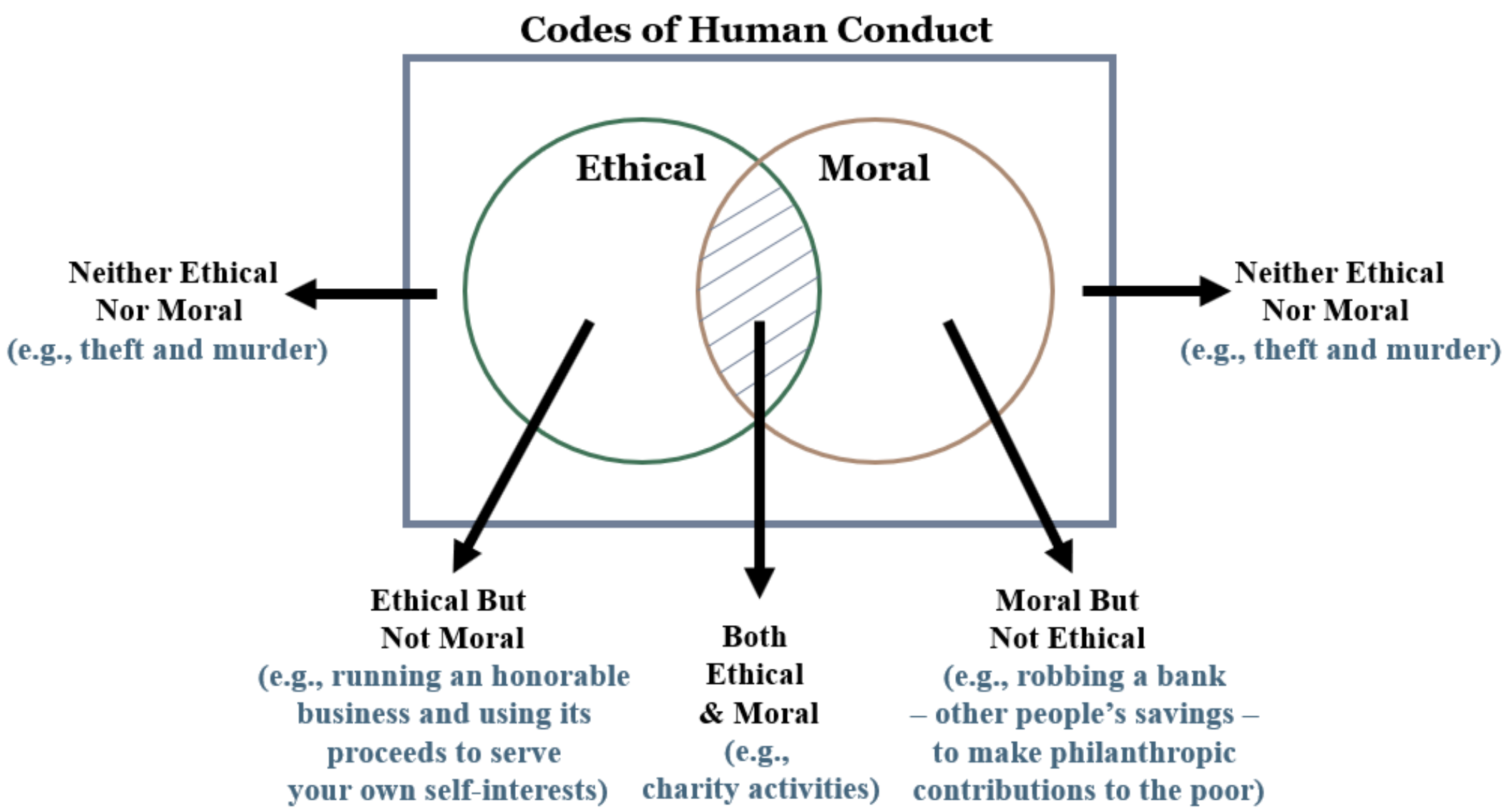

**Figure A.4.** Examples of Different Types of Codes of Human Conduct

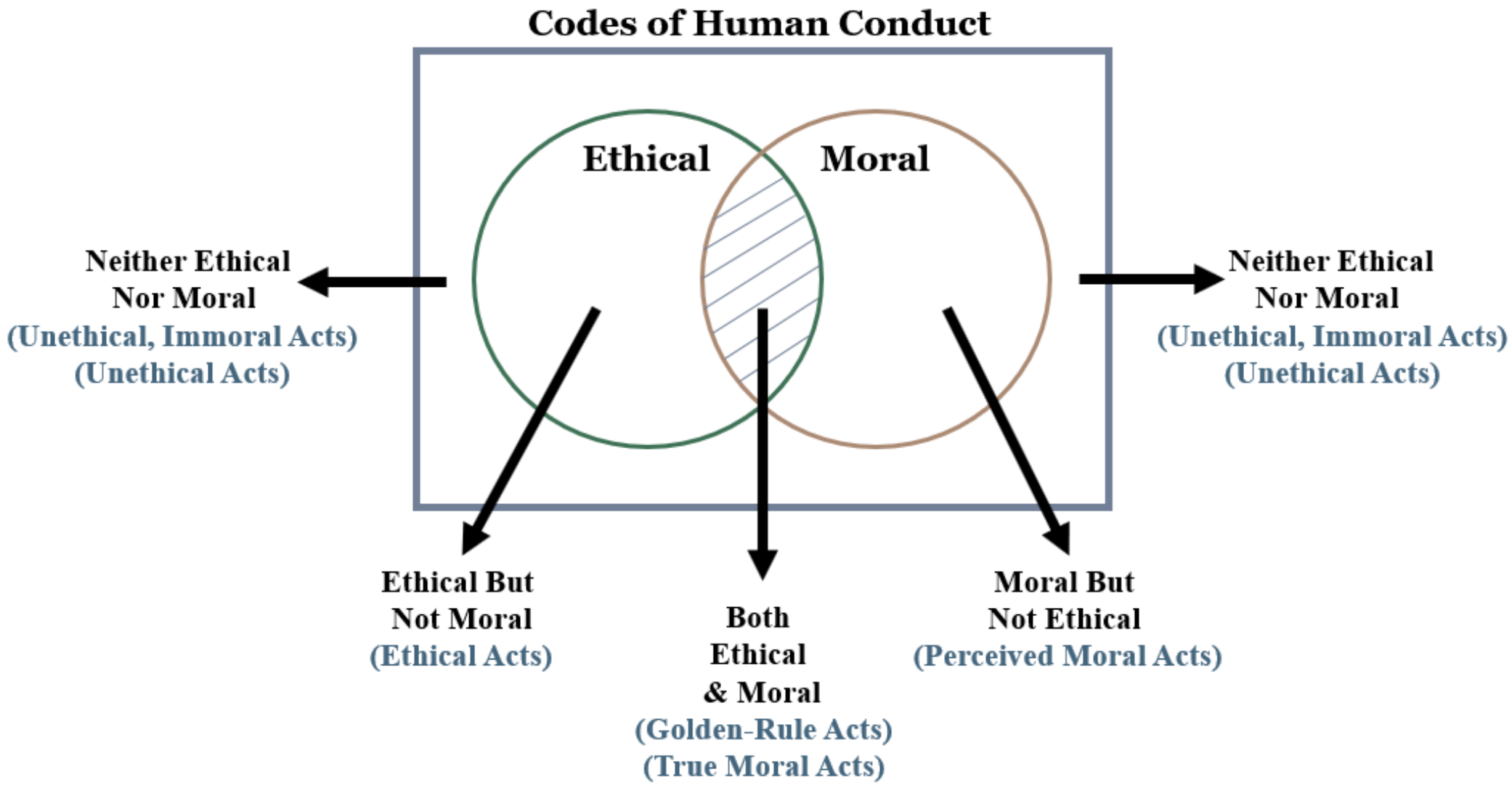

**Figure A.5.** Names of Acts Based on Different Types of Codes of Human Conduct

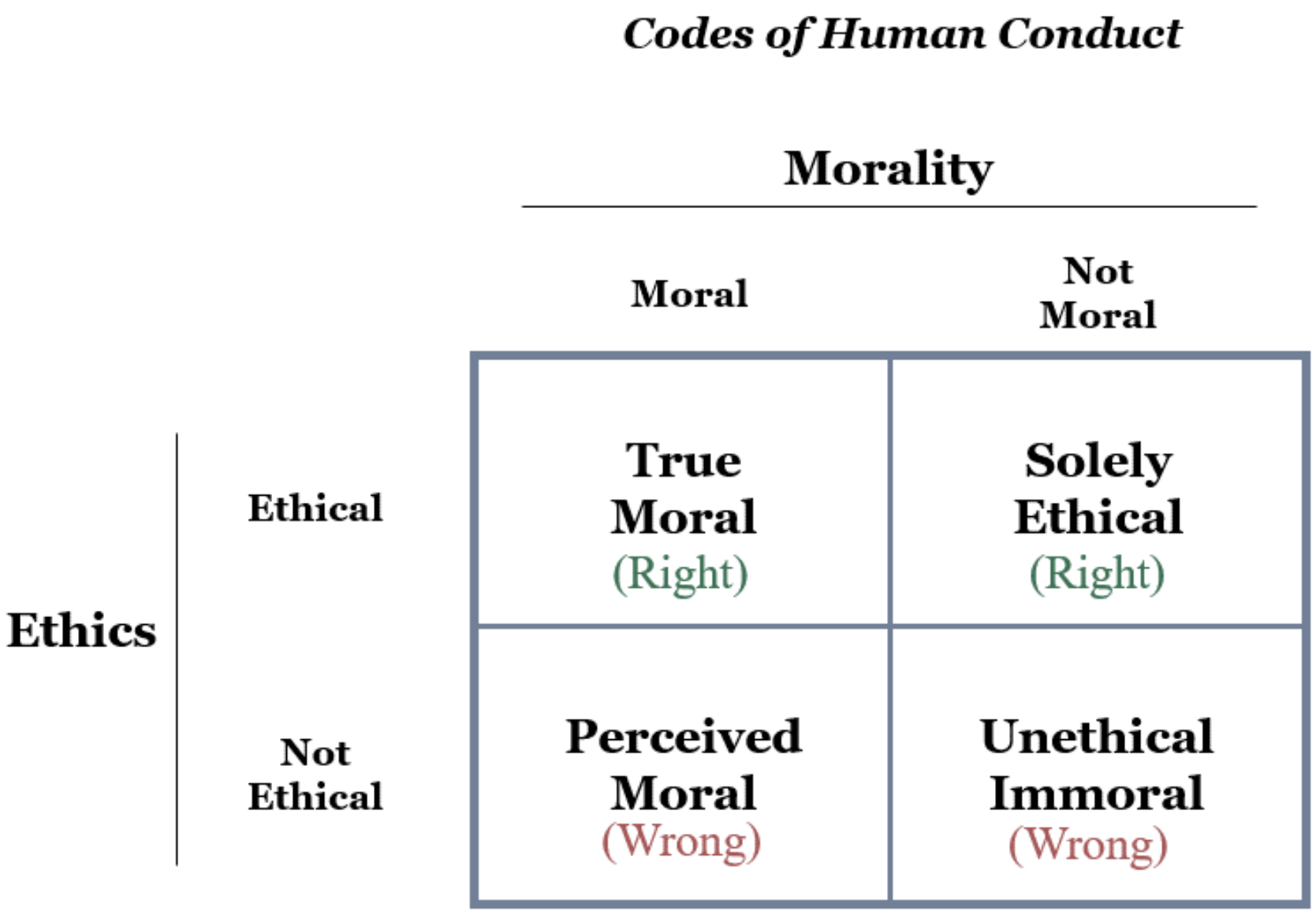

**Figure A.6.** Matrix of Acts Typology in Terms of Ethical and Moral Attributes